\documentclass[11pt]{article}
\usepackage{amsmath}
\usepackage{pstricks}
\usepackage[vcentermath]{youngtab}
\usepackage{bbm}
\usepackage{amssymb,amsfonts}

\newcommand{\re}[1] {(\ref{#1})}

\def\half{\frac{1}{2}}

\newfont{\bbbold}{msbm10 scaled \magstep1}

\def\bbC{\mbox{\bbbold C}}

\def\bbR{\mbox{\bbbold R}}

\def\bbZ{\mbox{\bbbold Z}}
\def\cA{{\cal A}}
\def\cB{{\cal B}}
\def\cC{{\cal C}}
\def\cD{{\cal D}}

\def\cG{{\cal G}}

\def\cJ{{\cal J}}

\def\cL{{\cal L}}

\def\cN{{\cal N}}
\def\cO{{\cal O}}

\def\cU{{\cal U}}

\newfont{\goth}{eufm10 scaled \magstep1}

\def\gg{\mbox{\goth g}}

\def\gl{\mbox{\goth l}}

\def\go{\mbox{\goth o}}
\def\gp{\mbox{\goth p}}

\def\gs{\mbox{\goth s}}

\def\gu{\mbox{\goth u}}

\def\a{\alpha}\def\adt{\dot \alpha}
\def\b{\beta}\def\bdt{\dot \beta}
\def\c{\gamma}\def\C{\Gamma}
\def\d{\delta}\def\D{\Delta}
\def\ve{\varepsilon}
\def\f{\phi}
\def\h{\eta}
\def\i{\iota}
\def\k{\kappa}
\def\l{\lambda}
\def\m{\mu}

\def\P{\Pi}
\def\r{\rho}
\def\s{\sigma}\def\S{\Sigma}

\def\th{\theta}

\def\be{\begin{equation}}\def\ee{\end{equation}}
\def\bea{\begin{eqnarray}}\def\eea{\end{eqnarray}}
\def\barr{\begin{array}}\def\earr{\end{array}}
\def\x{\xi}
\def\o{\omega}\def\O{\Omega}
\def\del{\partial}

\def\xz{\times}

\def\nab{\nabla}

\let\la=\label

\def\nn{\nonumber}
\def\bd{\begin{document}}
\def\ed{\end{document}}
\def\ba{\begin{array}}
\def\ea{\end{array}}
\def\bea{\begin{eqnarray}}
\def\eea{\end{eqnarray}}
\def\ft#1#2{\tfrac{#1}{#2}}
\def\fft#1#2{\frac{#1}{#2}}
\def\sst#1{{\scriptscriptstyle #1}}
\def\oneone{\rlap 1\mkern4mu{\rm l}}

\newcommand{\eq}[1]{(\ref{#1})}
\newcommand{\w}[1]{\\[0.#1cm]}
\def\eqs#1#2{(\ref{#1}-\ref{#2})}
\def\det{{\rm det\,}}
\def\tr{{\rm tr}}\def\str{{\rm str}}
\def\ad{{\rm ad}}

\newcommand{\hoch}[1]{$\, ^{#1}$}
\newcommand{\imperial}{\it\small Theoretical Physics Group, Imperial College London\\ Prince Consort Road, London SW7 2AZ, UK}
\newcommand{\kings}
{\it\small Department of Mathematics, King's College, University of London\\ Strand, London WC2R 2LS, UK}
\newcommand{\uu}
{\it\small Department of Theoretical Physics, Uppsala, Sweden}
\newcommand{\hip}
{\it\small HIP-Helsinki Institute of Physics, P.O. Box 64 FIN-00014
University of Helsinki, Suomi-Finland}
\newcommand{\stock}
{\it\small Department of Theoretical Physics, Stockholm, Sweden}
\newcommand{\golm}
{\it\small AEI, Max Planck Institut f\"ur Gravitationsphysik\\ Am M\"{u}hlenberg 1, D-14476 Potsdam, Germany}
\makeatletter
\renewcommand\theequation{\thesection.\arabic{equation}}
\@addtoreset{equation}{section} \makeatother

\newcommand{\sa}{/ \hspace{-1.2ex}}
\newcommand{\saa}{/ \hspace{-1.4ex}}
\newcommand{\saaa}{\, / \hspace{-1.6ex}}
\newcommand{\Scal}[1]{\Bigl ({#1} \Bigr )}
\newcommand{\scal}[1]{\bigl ({#1} \bigr )}

\newcommand{\CR}{\nonumber \\*}

\newcommand{\trace}{\hbox {tr}~}
\newcommand{\traceS}{\hbox {tr}_{\scriptscriptstyle \mathfrak{S}}~}

\DeclareMathAlphabet{\mathpzc}{OT1}{pzc}{m}{it}
\def\BRST{\,\mathpzc{s}\,}
\def\aBRST{{\scriptstyle (\mathpzc{s})}}
\def\q{{{\scriptscriptstyle (Q)}}}
\def\qs{{\scriptscriptstyle (Q\mathpzc{s})}}
\def\Qsla{{\mathcal{S}_{\q}}}
\def\Slav{{\mathcal{S}_\aBRST}}
\def\epsilonb{{\overline{\epsilon}}}
\def\bulletup{{\scriptstyle \bullet}}

\newcommand{\gra}[2]{{\scriptscriptstyle (#1 , #2 )}}
\newcommand{\ord}[1]{{\scriptscriptstyle (#1)}}

\def\cL{{\cal L}}
\def\cN{\mathcal{N}}
\def\cO{\mathcal{O}}

\def\ie{{\it i.e.}\ }
\def\eg{{\it e.g.}\ }

\newcommand{\sfrac}[2]{{\scriptstyle \frac{#1}{#2}}}
\newcommand{\stfrac}[2]{{\scriptscriptstyle \frac{#1}{#2}}}

 \def\balpha{{\overline{\alpha}}}
 \def\bbeta{{\overline{\beta}}}
 \def\bgamma{{\overline{\gamma}}}
 \def\bdelta{{\overline{\delta}}}
 \def\bepsilon{{\overline{\epsilon}}}
 \def\bvarepsilon{{\overline{\varepsilon}}}
 \def\bzeta{{\overline{\zeta}}}
 \def\bareta{{\overline{\eta}}}
 \def\btheta{{\overline{\theta}}}
 \def\bvartheta{{\overline{\vartheta}}}
 \def\biota{{\overline{\iota}}}
 \def\bkappa{{\overline{\kappa}}}
 \def\blambda{{\overline{\lambda}}}
 \def\bmu{{\overline{\mu}}}
 \def\bnu{{\overline{\nu}}}
 \def\bxi{{\overline{\xi}}}
 \def\bpi{{\overline{\pi}}}
 \def\brho{{\overline{\rho}}}
 \def\bvarrho{{\overline{\varrho}}}
 \def\bsigma{{\overline{\sigma}}}
 \def\bvarsigma{{\overline{\varsigma}}}
 \def\btau{{\overline{\tau}}}
 \def\bphi{{\overline{\phi}}}
 \def\bvarphi{{\overline{\varphi}}}
 \def\bchi{{\overline{\chi}}}
 \def\bpsi{{\overline{\psi}}}
 \def\bomega{{\overline{\omega}}}

\def\thalf{{\textrm{\tiny\textonehalf}}}
\def\tquarter{{\textrm{\tiny\textonequarter}}}
\def\Ko{{\scriptscriptstyle K}}
\def\tKo{\scriptscriptstyle k }
\def\corr{$\clubsuit$}

\newcommand{\auth}{\large P.S.\ Howe${}^{a,}$\footnote{email: paul.howe@kcl.ac.uk} and U. Lindstr\"om${}^{b,c,}$\footnote{email: ulf.lindstrom@physics.uu.se}}

\begin{document}

\renewcommand{\thefootnote}{\fnsymbol{footnote}}

\null
\begin{flushright}
{\small KCL-MTH-15-07}\\
{\small UUITP-16/15}\\
{\small Imperial-TP-UL-2015-02}\\
\vskip 1.5 cm
\end{flushright}

\begin{center}
{\Large{\bf Notes on Super Killing Tensors}}
\vspace{.75cm}

\auth
\end{center}
\vspace{.5cm}

\centerline{${}^a${\it \small Department of Mathematics, King's College London}}
\centerline{{\it \small The Strand, London WC2R 2LS, UK}}
\vspace{.5cm}
\centerline{${}^b${\it \small Department of Physics and Astronomy, Theoretical Physics, Uppsala University}}
\centerline{{\it \small SE-751 20 Uppsala, Sweden }}
\vspace{.5cm}
\centerline{${}^c${\it \small Theoretical Physics, Imperial College, London}}
\centerline{{\it \small Prince Consort Road, London SW7 2AZ, UK}}

\vspace{1cm}


\centerline{{\bf Abstract}}
\vskip .5cm
The notion of a Killing tensor is generalised to a superspace setting. Conserved quantities associated with these are defined for superparticles and Poisson brackets are used to define a supersymmetric version of the even Schouten-Nijenhuis bracket. Superconformal Killing tensors in flat superspaces are studied for spacetime dimensions 3,4,5,6 and 10. These tensors are also presented in analytic superspaces and super-twistor spaces for 3,4 and 6 dimensions. Algebraic structures associated with superconformal Killing tensors are also briefly discussed.

\vspace{1cm}


\renewcommand{\thefootnote}{\arabic{footnote}}
\setcounter{footnote}{0}

\pagebreak
\tableofcontents
\setcounter{page}{1}


\section{Introduction}

In ordinary Lorentzian spacetime a Killing vector, $K$, is a symmetry of the metric,  \ie a vector field that generates an infinitesimal diffeomorphism that leaves the metric $g$ invariant, $\mathcal{L}_K g=0$.  With respect to an orthonormal basis related to a coordinate basis by the vielbein $e_m{}^a$ a Killing vector (KV) satisfies
\be
\nab_a K^b=L_a{}^b\ ,
\la{1.1}
\ee
where $L_{ab}=-L_{ba}$ and  $\nab$ is a metric covariant derivative; for a conformal Killing vector (CKV) we have
\be
\nab_a K^b=\tilde L_a{}^b\ ,
\la{1.2}
\ee
where $\tilde L_{ab}=L_{ab} + 2\h_{ab} S$, where $\h_{ab}$ denotes the standard flat components of the metric in an orthonormal frame. In other words a Killing vector is constant up to a Lorentz transformation while a conformal Killing vector is constant up to a Lorentz transformation together with a scale transformation. We note also that, in the Hamiltonian formalism for a massless particle, a CKV is a function on phase space linear in the momentum whose Poisson bracket with the Hamiltonian vanishes weakly.

These relations can readily be generalised to higher-order Killing tensors (KTs). These are symmetric tensors obeying
\be
\nab_a K^{b_1\ldots b_n}=L_a{}^{(b_1,b_2\dots b_n)}\ ,
\la{1.3}
\ee
where $L_{ab_1,b_2\ldots b_n}$ is antisymmetric on the first two indices and symmetric on the rest. An $n$th rank conformal Killing tensor (CKT) is symmetric, traceless and obeys
\be
\nab_a K^{b_1\ldots b_n}=\tilde L_a{}^{\{b_1,b_2\dots b_n\}}\ ,
\la{1.4}
\ee
where $\tilde L_{ab_1,b_2\ldots b_n}$ now includes a trace part on the first two indices and where the curly brackets denote traceless symmetrisation. In the context of massless particles moving along geodesics in the given spacetime, conformal Killing tensors lead to higher-order constants of the motion defined by
\be
K=K^{a_1\ldots a_n} p_{a_1}\ldots p_{a_n}\ ,
\la{1.5}
\ee
where the momentum $p$ is covariantly constant with respect to suitable time parameter. It is straightforward to see that the (traceless) symmetrised product of two (conformal) Killing vectors is a second-rank (conformal) Killing tensor, so that any spacetime that admits the former will also have (conformal) Killing tensors\footnote{The opposite is not true, there are cases where there are (irreducible) Killing tensors but no Killing vectors \cite{Dolan2}.}.

In this article we shall generalise the basic concepts discussed above to superspace, building on earlier discussions of various aspects of superconformal symmetry in superspace, see, for example, \cite{Sohnius:1976pa,Ferber:1977qx,Lang:1981dp,Bonora:1984pn,Shizuya:1986xt,Howe:1995md,Buchbinder:1998qv}.\footnote{In the mathematics literature one can find supersymmetric versions of Killing vectors defined on super-Riemannian spaces equipped with ortho-symplectic metrics \cite{Coul:2012}.  In this paper we shall focus on superspaces as commonly understood in physics. These extend spacetime by sets of odd spinorial coordinates.} This is not entirely straightforward due to the presence of constraints in supergeometries. Having discussed the general case we shall apply it to  conservation laws for superparticles where we use a closed two-form on the even tangent bundle to define Poisson brackets for these conserved quantities. This turns out to be well-defined even though the two-form is itself singular. Given this Poisson bracket structure we can derive a bracket for superconformal Killing tensors that extends the even Schouten-Nijenhuis bracket (see below) to the super case. We then turn to a discussion of superconformal Killing tensors in flat superspaces, specifically in dimensions 3,4,6,5 and 10. In the first three cases there are classical superconformal groups, in $D=5$ there is an exceptional superconformal group, $F(4)$, only for the case $N=1$ while in ten dimensions the compensating scale parameter is required to be constant \cite{Nahm:1977tg}. We also discuss the first three cases in analytic superspaces. These are particular coset superspaces of the superconformal group for which the local description resembles spacetime considered in a similar way. 

In flat superspaces superconformal Killing tensors (SCKTs) are given by finite-dimensional representations of the corresponding Lie superalgebras \cite{Scheunert:1976uf,Scheunert:1976ug,Kac:1977qb,Kac:1977em}. One feature that is not present in the purely even case is that some representations can be reducible but indecomposable, see \cite{Bars:1982se,Leites:2002} and references therein. We discuss examples of this in super Minkowski space, analytic superspace and also in super-twistor spaces in sections 4,5, and 6 respectively. In particular, it turns out that the definition of SCKTs given in \eq{2.12} below  is not always sufficient, and furthermore, that when appropriate additional constraints are imposed there are invariances that correspond to the presence of sub-representations that cannot be removed due to indecomposablity. A familiar example of this is given by a SCKV in $N=4, D=4$ supersymmetry where one has to impose super-tracelessness separately and where one is still left with an additional one-parameter symmetry that reduces the algebra to $\gp\gs\gl(4|4)$. In section 6 we also comment on the possibility of defining algebras associated with SCKTs in a similar way to those that arise as symmetries of the Laplacian in the purely even case \cite{Eastwood:2002su}. Such algebras may have applications in higher-spin theory as we briefly comment on in section 7.  We end in section 8 with a few concluding remarks.

\section{Killing tensors in superspace}

We consider superspaces that extend $D$-dimensional spacetime by a number of odd coordinates that transform as spinors under $Spin(1,D-1)$ and which, in addition, may carry a representation of an internal R-symmetry group. Superspace was introduced in \cite{Salam:1974yz,Ferrara:1974ac} and generalised to the curved case in \cite{Ogievetsky:1976qb,Wess:1977fn,Ogievetsky:1978mt,Siegel:1978mj}.\footnote{See also \cite{Akulov:1974xz} where superspace was introduced in the context of non-linear realisations of supersymmetry.}
We denote the superspace coordinates by $z^M=(x^m,\th^\m)$. There is no super-metric but there is a super-vielbein $E_M{}^A$ that relates coordinate and preferred frame bases by $E^A= dz^M E_M{}^A$. The basic structure is a choice of odd tangent bundle $T_1\subset T$, the tangent bundle, such that $T_1$ generates the even tangent bundle $T_0=T/T_1$ by Lie brackets \cite{Siegel:1978mj}; in other words, $T_1$ is maximally non-integrable. In addition, we suppose that $T_1=S\otimes V$ where $S$ is a spinor bundle and $V$ a bundle carrying the fundamental representation of the internal R-symmetry group.\footnote{See, for example, \cite{Manin:1988ds}, for a more formal discussion.} This then reduces the structure group to a triangular form but it is standard practice to reduce it further to a diagonal one by an appropriate choice of $T_0$, and we shall always make such a choice in what follows. Given such a choice, the tangent bundle splits into even and odd so that the structure group does not mix them. Thus we have $E^A=(E^a, E^\a)$. The structure group is taken to be the Lorentz group acting on even (vector) indices $a,b$ etc and the product of the corresponding spin group and any R-symmetry group acting on odd (spinorial) indices $\a,\b$ etc. We use a two-step notation for the odd indices. We let $\a$ run over all the odd indices, but when necessary, we shall replace $\a$ by a pair $\a i$, where now $\a$ runs over the dimension of the appropriate spin representation and $i$ is an R-symmetry index, corresponding to $S$ and $V$ respectively.  We then introduce a connection one-form $\O_A{}^B$ taking its values in the Lie algebra of the structure group (so that there are no mixed components), and correspondingly define the torsion and curvature forms in the usual way: $T^A=D E^A:=d E^A+ E^B \O_B{}^A$ and $R_A{}^B=d \O_A{}^B + \O_A{}^C \O_C{}^B$. The various components of the connection can be determined in terms of the vielbein if we impose suitable conventional constraints on the torsion. In addition, we can impose further constraints that specify some parts of the super-vielbein. Finally, throughout this paper, we shall assume that the dimension-zero torsion is flat,
\be
T_{\a\b}{}^c=-i(\C^c)_{\a\b}\ ,
\la{2.1}
\ee
where $\C^c$ denotes a product of the appropriate gamma-matrix and an R-invariant tensor, if needed ($\C^c$ is symmetric on the joint indices $\a,\b$).  The dimension-zero torsion takes this form in supergravity, when the equations of motion are satisfied, and sometimes off-shell. In the presence of higher-order string or M-theory corrections it may be that this is not the case, for example in $D=11$ supergravity \cite{Howe:1997he,Cederwall:2004cg}, but we shall not consider this possibility here; we shall always assume that \eq{2.1} holds. In addition we shall impose some conventional constraints that do not depend on the spacetime-dimension-dependent nature of the spinors. At dimension one-half we can take  
\be
T_{\a [bc]}=0 \qquad {\rm and } \qquad (\C_c)^{\a\b} T_{\b b}{}^c=0\ .
\la{2.1a}
\ee
The first of these allows one to solve for the dimension one-half component of the Lorentz connection, while the second allows one to fix the splitting of the tangent space into odd and even. These two constraints imply that the remaining component of this torsion is symmetric, traceless and gamma-traceless. This is an irreducible representation  of the spin group that is not present in the other dimension-one-half torsion $T_{\a\b}{}^\c$. It  then  follows from the dimension-one-half Bianchi identity that this must also be zero so that \eq{2.1a} implies
\be
T_{\a b}{}^c=0\ .
\la{2.1b}
\ee
We can also choose
\be
T_{ab}{}^c=0
\la{2.1c}
\ee
as a conventional constraint for the Lorentz connection at dimension one. In addition, we shall assume that conventional constraints corresponding to the dimension one-half and one components of the R-symmetry connection have been imposed, but  it will not be necessary to be explicit about these in this paper.

The natural generalisation of a Killing vector to superspace would seem to be a vector field $K^A$ that satisfies\footnote{For other discussions of this topic in curved superspace see \cite{Buchbinder:1998qv,Kuzenko:2015lca}.}
\be
\nab_A{}K^B + K^C T_{CA}{}^B=L_A{}^B
\la{2.2}
\ee
where $L_A{}^B$ denotes an element of the Lie algebra of the structure group.  However, as we shall see, the constraints that we have imposed on the geometry mean that the full vector field is determined by its even part, $K^a$, from the lowest-dimensional component (\ie dimension minus one-half) of \eq{2.2}, namely
\be
\nab_\a K^b -i K^\c (\C^b)_{\c\a}=0\ ,
\la{2.4}
\ee
when \eq{2.1} and \eq{2.1b} are imposed. The spinorial derivative $\nab_\a$ acting on $K^b$ gives two representations of the spin group, a gamma-traceless vector-spinor and  a spinor, and \eq{2.4} states that the former should vanish. The spinor is then determined in terms of the spinorial derivative of $K^b$. The vector field $K^A=(K^a,K^\a)$, where the spinorial component is determined in this fashion, then satisfies \eq{2.2} with $\tilde L_A{}^B$ on the right, where the tilde indicates the inclusion of an appropriate super-Weyl transformation \cite{Siegel:1978nn,Howe:1978km}.   In other words, \eq{2.4} defines a superconformal Killing vector (SCKV) when the standard constraints given above are imposed.

To see this  directly, we apply a second odd covariant derivative to both sides of \eq{2.4}, take the graded commutator and then make use of the first Bianchi identity, $DT^A=E^B R_B{}^A$, to get (when \eq{2.1} holds), 
\be
(\C^a)_{\a\b}(\nab_a K^b + K^C T_{Ca}{}^b)=(\nab_{\a} K^\c + K^D T_{D \a}{}^\c) (\C^b)_{\c\b}\  + \ (\a \leftrightarrow\b)\ .
\la{2.5}
\ee
This equation tells us that the dimension-zero components of \eq{2.2} (\ie those for which the indices $(A,B)$ are either both even or both odd) are satisfied if
\be
\tilde L_a{}^b=L_a{}^b + 2\d_a{}^b S\qquad \tilde L_\a{}^\b=L_\a{}^\b + \d_\a{}^\b S\ ,
\la{2.6}
\ee
where $S$ is a local scale parameter. Explicitly,
\begin{align}
\nab_a K^b &=\tilde L_a{}^b\, =L_a{}^b + 2\d_a{}^b S   \label{2.6.1}\w1
\nab_{\a} K^\c + K^D T_{D \a}{}^\c&=\tilde L_\a{}^\b=L_\a{}^\b + \d_\a{}^\b S\ ,
\label{2.6.2}
\end{align}
where, in \eq{2.6.1}, we have used \eq{2.1b} and \eq{2.1c}. This equation is now formally identical to \eq{1.2}.
We can then use a similar argument to show that, at dimension one-half,
\be
\nab_a{}K^\b+ K^D T_{D a}{}^\b=iA (\C_a)^{\b\c} \nab_{\c} S\ ,
\la{2.7}
\ee
where $A$ is a constant. Depending on the theory (\ie on any additional constraints that have been imposed) it might be the case that $\nab_\a S=0$, in which the scale parameter is a constant. 

Another way of deriving this result is to consider  arbitrary variations of the supervielbein, $H_A{}^B:=E_A{}^M \d E_M{}^B$, and connection,
$\f_{A,B}{}^C:=E_A{}^M\d \O_{M,A}{}^B$ (where $E_A{}^M$ is the inverse supervielbein). The induced variation of the torsion is
\be
\d T_{AB}{}^C=2 \nab_{[A} H_{B]}{}^C + T_{AB}{}^D H_D{}^C -2H_{[A}{}^D T_{|D| B]}{}^C + 2 \f_{[A,B]}{}^C\ ,
\la{2.8}
\ee
where the square brackets denote graded antisymmetrisation.\footnote{By graded (anti)-symmetrisation of indices we shall mean $\bbZ_2$-grading throughout the paper.} For a diffeomorphism accompanied by a local structure-algebra transformation we have
\bea
H_A{}^B&=&\nab_A K^B + K^C T_{CA}{}^B - L_A{}^B\nn\w1
\f_{A,B}{}^C&=&\nab_A L_B{}^C + K^D R_{DA,B}{}^C\nn\w1
\d T_{AB}{}^C&=& K^D\nab_D T_{AB}{}^C + 2L_{[A}{}^D T_{|D|B]}{}^C-T_{AB}{}^D L_D{}^C\ .
\la{2.9}
\eea
For a Killing symmetry the variation of the supervielbein must vanish up to a local frame rotation so that $H_A{}^B$=0.  Since $\f_{A,B}{}^C$ is determined in terms of $H_A{}^B$ (by setting appropriate parts of the torsion to zero), the variation of the torsion must also vanish. Now suppose that we make a diffeomorphism but only set $H_\a{}^b=0$. The dimension-zero component of \eq{2.8} reduces to 
\be
0=(\C^c)_{\a\b} H_d{}^c -2H_{(\a}{}^\c (\C^c)_{\b)\c}\ .
\la{2.10}
\ee
This clearly implies that the dimension-zero components of $H_A{}^B$ will also vanish up to a scale transformation since a structure-algebra transformation preserves $\C^c$. Now consider the variation of the dimension-one-half torsion component $T_{\a b,c}$. It is easy to see that setting $T_{\a bc}=0$ allows one to solve for the Lorentz part of the dimension-one-half connection variation $\f_{a,bc}$, as well as  $H_a{}^\b$. But since the only other non-zero terms in the equation contributing to this component involves the derivative of $S$, we conclude that $H_a{}^\b$ indeed has the same form as the right-hand side of \eq{2.7} and thus that we have a superconformal Killing vector.

This result means that we can interpret a SCKV in a simpler way:  it can be defined to be a vector field K that generates an infinitesimal diffeomorphism that preserves the odd tangent bundle, \ie $\mathcal{L}_K X$ is odd if $X$ is, or, equivalently
\be
<[E_\a,K],E^b>=0\ .
\la{2.11}
\ee
Here $E_A$ denote basis vector fields dual to the basis one-forms $E^A$ and the angle-brackets denote the standard pairing between forms and vectors. Writing this out explicitly gives \eq{2.4}, and this leads to \eq{2.6.1}, \eq{2.6.2} and \eq{2.7}, as shown above.

The above results can be extended to superconformal Killing tensors (SCKTs) straightforwardly. Thus a SCKT  is determined by a symmetric traceless purely even tensor $K^{b_1\ldots b_n}$. Its covariant spinorial derivative again contains just two irreducible spinorial representations of the Lorentz group, one with $n$ vector indices, and one with $(n-1)$, both of which are symmetric-traceless on these and gamma-traceless. To obtain a SCKT we simply have to set the larger representation to zero. Thus we have
\be
\nab_\a K^{b_1\ldots b_n} -i nK^{\{b_1\ldots b_{n-1} \c}( \C^{b_n\}})_{\c\a}{}=0\ 
\la{2.12}
\ee
when the standard constraints are satisfied. Note that we can take $K^{b_1\ldots b_{n-1} \c}$ to be irreducible because any gamma-trace term it could contain drops out of \eq{2.12}.
Given this one would then expect to be able to construct all the other components of a full SCKT $K^{A_1\ldots A_n}$ systematically by applying further spinorial covariant derivatives to \eq{2.12} and making use of the Ricci and Bianchi identities. One would expect that this object should satisfy
\be
\nab_A K^{B_1\ldots B_n} + n K^{\{B_1\ldots B_{n-1} C} T_{CA}{}^{B_n\}}=\tilde L_A{}^{\{B_1,B_2\ldots B_n\}}\ 
\la{2.3}
\ee
for some appropriate definition of the brackets $\{\}$, although is not straightforward to verify this explicitly in the general case.
Later on, in section 4, we shall work out all of the components of SCKTs in flat superspaces. The emphasis throughout the rest of the paper will be on the superconformal case, but  the non-conformal case can be studied when the scale transformations are omitted.

\section{Superparticles}

The Lagrangian for a Brink-Schwarz superparticle in a curved background is given by  \cite{Brink:1981nb,Siegel:1983hh}
\be
L=\half e^{-1} \dot z^a \dot z_a
\la{3.1}
\ee
Here the coordinates of the superparticle moving along a curve parametrised by $t$ are given by $z^M(t)$, and we set $\dot z^A:=\dot z^M E_M{}^A$.  The equations of motion are
\begin{align}
\nab_t p_a + \dot z^C T_{Ca}{}^b p_b&=0\ ,\nn\w1
\dot z^B T_{B\a}{}^c p_c&=0\ ,
\la{3.2}
\end{align}
where $p_a=e^{-1} \dot z_a$ is  the derivative of the Lagrangian with respect to $\dot z^a$. Varying the action with respect to the einbein $e$ gives the mass-shell constraint $p^2=0$. The covariant derivative $\nab_t$ is the pull-back of the superspace covariant derivative onto the worldline. From now on we suppose that the standard superspace constraints given by \eq{2.1}, \eq{2.1a}, \eq{2.1b} and \eq{2.1c}, are satisfied.  The equations of motion simplify to
\begin{align}
\nab_t p_a &=0\ ,\nn\w1
\dot z^\b (\C\cdot p)_{\b\a}&=0\ .
\la{3.2a}
\end{align}
The superparticle action is invariant under the (fermionic) kappa-symmetry transformations introduced in \cite{Siegel:1983hh}:
\begin{align}
\d z^\a&=(\C\cdot p)^{\a\b} \k_\b\ ,\nn\w1
\d e&=-2i\h \dot z^\a \k_\a\ ,
\la{3.2b}
\end{align}
where $\h=\pm 1$ depending on the dimension.

Now consider the function $K=K^{a_1\ldots a_n}p_{a_1}\ldots p_{a_n}$, where $K$ is an $n$th rank symmetric, traceless tensor. We claim that this is conserved along the worldline of the superparticle when the equations of motion are satisfied provided that this tensor satisfies \eq{2.12}. We have
\be
\frac{d K}{dt}=(\dot z^b \nab_b K^{a_1\ldots a_n} + \dot z^\b \nab_\b K^{a_1\ldots a_n})p_{a_1}\ldots p_{a_n}\ ,
\la{3.2c}
\ee
where we have used the fact that $\nab_t p_a=0$ along the worldline. For this expression to be zero the two terms have to vanish independently. In the second term on the right $\nab_\b K^{a_1\ldots a_n}$ contains only two irreducible representations because $K^{a_1\ldots a_n}$ is symmetric and traceless. These are an $n$th rank symmetric, traceless, gamma-traceless tensor-spinor and a similar object with $(n-1)$ tensor indices. Clearly the former will not give zero in the above equation and hence must be set to zero in order for it to be true. Thus the second term gives a zero contribution  if and only if \eq{2.12} holds. When it does then the first term in \eq{3.2c} vanishes as well: since $\dot z^a\propto p^a$, the first term involves $\nab^{\{b} K^{a_1\ldots a_n\}}$ on-shell, and this vanishes for a SCKT. So we have the result that the function $K$ is conserved if and only if \eq{2.12} holds, \ie if $K^{a_1\ldots a_n}$ defines a SCKT.

We can also show that the function $K$ is invariant under kappa-symmetry. We have
\begin{align}
\d_\k K&= \d_\k z^\a \nab_\a K^{b_1\ldots b_n} p_{b_1}\ldots p_{b_n}\nn\w1
&= in (\C\cdot p)^{\a\b} \k_\b (\C^{b_1})_{\a\c} K^{b_2\ldots b_n\c}p_{b_1}\ldots p_{b_n}\nn\w1
&=i\h n(\C^{b_1} \C^a\k)_\c K^{b_2\ldots b_n\c}p_ap_{b_1}\ldots p_{b_n} =0\qquad {\rm on\  shell}\ .
\la{3.2d}
\end{align}
The variation of the momentum can be ignored as it vanishes on-shell,
\be
\d_\k p_a=i\h \dot z^\a(\C^b p_b \C_a \k)_\a=0\  ,
\la{3.2e}
\ee
by virtue of the equation of motion for $\dot z^\a$. This is consistent as kappa-symmetry can be thought of as extended world-line supersymmetry in the super-embedding approach  \cite{Sorokin:1988nj,Sorokin:1989zi}, so that the commutator of two such transformations gives rise to a time translation.

We shall now discuss the supersymmetric extension of the even Schouten-Nijenhuis bracket for CKTs. We briefly review the bosonic case. For a spinless zero-mass particle the Lagrangian is $L=\half e^{-1} \dot x^a \dot x_a$ where $\dot x^a=\dot x^m e_m{}^a$, where $e_m{}^a$ is the vielbein, the momentum is $p_a=\frac{\del L}{\del \dot x^a}=e^{-1}\dot x_a$ while the momentum associated with the einbein, $p_e$, is zero. The Hamiltonian is $H=\half e p^2$. The Poisson bracket of the constraint $p_e\cong 0$ with the Hamiltonian is $-H$ so that the Hamiltonian is weakly zero. The symplectic form on phase space is
\be
\s=e^a Dp_a:=e^a (dp_a-\o_a{}^b p_b)\ ,
\ee
where $\o_a{}^b$ is the standard torsion-free connection one-form and $e^a=dx^m e_m{}^a$. The basis of vector fields dual to $(e^a,Dp_a)$ is $(\tilde e_a,\del^a)$ where
\be
\tilde e_a:=e_a{}^m(\del_m +\o_{m,b}{}^c p_c \del^b) \qquad \del^a:=\frac{\del}{\del p_a}\ .
\ee
A Hamiltonian vector field $X_f$, corresponding to a function $f$, satisfies
\be
\i_{X_f}\s= d f\,
\la{3.9}
\ee
explicitly
\be
X_f=(\tilde e_a f )\del^a-(\del^a f) \tilde e_a\ .
\la{3.9a}
\ee
We define the Poisson bracket of two functions by
\be
(f,g)=-\i_{X_g}\i_{X_f} \s \ .
\la{3.10}
\ee
With respect to a covariant basis is easily seen to be
\be
(f,g)=\tilde e_a f \del^a g- f\leftrightarrow g\ ,
\la{3.7}
\ee

A CKT can now be defined as a function on phase space whose Poisson bracket with the Hamiltonian vanishes weakly. 
Writing such a function as $K:=K^{a_1\ldots a_n}p_{a_1}\ldots p_{a_n}$, it is easy to see that this constraint is precisely \eq{1.4}. If $L$ is another such function, with a rank $m$ symmetric traceless tensor, then the fact that the Poisson bracket obeys the Jacobi identity implies that $(K,L)$ also has a weakly vanishing Poisson bracket with the Hamiltonian. We can write (minus) this as a new tensor constructed from $K$ and $L$ multiplied by $(n+m-1)$ factors of the momentum.\footnote{The minus sign is so that $[K,L]^a$ for two vectors is the Lie bracket.} This new tensor defines the even Schouten-Nijenhuis bracket $[K,L]$,
\be
[K,L]^{a_1\ldots a_q}:=nK^{\{a_1\ldots a_{n-1}|b|}\nab_b L^{a_n\ldots a_q\}}-mL^{\{a_1\ldots a_{m-1}|b|}\nab_b K^{a_m\ldots a_q\}} \ ,
\la{3.8}
\ee
 where $q=m+n-1$. In a coordinate basis the covariant derivative $\nab_b$ can be replaced by an ordinary partial derivative. Note that, in the literature, there are two brackets attributed to these authors \cite{Sch,Nij}. The one we refer to here as {\em even} involves symmetric contravariant tensors, while the other, which  could be called {\em odd}, involves anti-symmetric contravariant tensors (or multivectors). Henceforth we shall only be interested in the even bracket which we shall refer to as the SN bracket. This is the term used for such brackets in the physics literature in the context of Killing tensors \cite{Dolan}.
The even bracket, as noted above, is related to the Poisson bracket on the contangent bundle $T^*M$, while the odd bracket can be derived in a similar way from the anti-bracket defined from the anti-symplectic two-form on the Grassmann-flipped cotangent bundle $\P T^*M$ (i.e. the fibre coordinates are taken to be odd). Discussions of this topic and other variations can be found in, for example, \cite{DuboisViolette:1994gy,Soroka2}.\footnote{In the context of particles it is also possible to have supersymmetry on the worldline rather than the ambient space; this is called a spinning particle. For discussions of generalised Killing tensors in this context, see, for example, \cite{Gibbons:1993ap,Santillan:2011sh}.}

The foregoing can be extended to the superparticle case although not quite straightforwardly. In the super case the phase space is the even cotangent bundle coordinatised by $x^m,\th^\m$ and $p_a$. There is a natural closed two-form $\S$ \cite{Howe:1991hk} given by 
\be
\S:=E^a Dp_a + T^a p_a\ .
\la{3.9.1}
\ee
although it is not symplectic because it is singular on the mass-shell  ($p^2=0$) as will become apparent below. However, we can still define Hamiltonian vector fields as before, with $\s$ replaced by $\S$, and we find
\be
X_f=(\tilde E_a f)\del^a  - (\del^a f)\tilde E_a +( i\h\frac{(\C\cdot p)^{\a\b}}{p^2}\tilde E_\b f)\tilde E_\a\  ,
\la{3.11}
\ee
where $\tilde E_A=E_A{}^M(\del_M+\O_{M,b}{}^c p_c\del^b)$ and where $\h=\pm 1$ depending on the dimension. The Poisson bracket is
\be
(f,g)=(\tilde E_a f\del^a g -f\leftrightarrow g) +i\h \tilde E_\a f  \frac{(\C\cdot p)^{\a\b}}{p^2} \tilde E_\b g.
\la{3.12}
\ee
It is clear from these formulae that $\S$ is not invertible on-shell, but this singularity cancels out when we compute the Poisson bracket of two conserved functions defined by SCFTs.

We cannot repeat the arguments given for symmetries of the bosonic particle straightforwardly because the phase space does not include the fermionic momenta, and, as is well-known, the constraint structure of the superparticle does not allow a simple covariant discussion. If we start with a function $K$ of the same form as in the bosonic case, with the difference that now $K^{a_1\ldots a_n}$ depends on $\th$ as well as $x$, then, demanding that the Poisson bracket of $K$ with the Hamiltonian $H$ be weakly zero leads to the constraint $\nab^{\{a_1} K^{a_2\ldots a_{(n+1)}\}}=0$, which is formally the same as the bosonic case. This is a consequence of \eq{2.12} but does not imply it. We are therefore obliged to impose \eq{2.12} as a constraint. If we compute the Poisson bracket of two such functions, $K$ and $L$, defined by symmetric, traceless tensors with $n$ and $m$ indices respectively, the $p^2$ in the denominator of the bracket cancels so we find a new function with an $(n+m-1)$th-rank tensor given by the supersymmetric (even) Schouten-Nijenhuis bracket.

\begin{align}
[K,L]^{a_1\ldots a_q}&:=nK^{\{a_1\ldots a_{n-1}|b|}\nab_b L^{a_n\ldots a_q\}}-mL^{\{a_1\ldots a_{m-1}|b|}\nab_b K^{a_m\ldots a_q\}}\nn\w1
&\ \ \ \   +imnK^{\{a_1\ldots a_{n-1} \c}(\C^{a_n})_{\c\d} L^{a_{n+1}\ldots a_q\}\d}\ .
\label{3.13}
\end{align}

We emphasise that it is {\em not} guaranteed that this tensor also satisfies \eq{2.12}, but it can be verified directly that it does, as shown in the appendix. We therefore conclude that the supersymmetric SN bracket defined above for SCKTs does indeed define a new SCKT. Moreover, since the Poisson bracket obeys the Jacobi identity, so does the super-SN bracket and thus we have a Lie algebra structure on the space of SCKTs.

\section{SCKTs in flat superspaces}

In this section we give the details of SCKTs in flat superspaces in $D=3,4,6,5\,\&10$. In the first three cases we shall consider arbitrary numbers ($N$) of supersymmetries, while in $D=5$ there is only one case, $N=1$, for which one can have superconformal transformations \cite{Nahm:1977tg}. In $D=10$, where there are no conformal boosts or $S$-supersymmetry transformations, we shall only consider $N=1$.  In $D=3,4,6$ there are superconformal groups, $SpO(2|N), SU(2,2|N)$ and $OSp(8|N)$, and the SCKVs represent the corresponding Lie superalgebras, while in $D=5, N=1$, the superconformal group is the exceptional Lie supergroup $F(4)$.\footnote{$SpO(2|N)$ is the same as $OSp(N|2)$ but we write it with the symplectic factor first in $D=3$ to emphasise that this refers to the spacetime conformal  group.}
In $D=10$ on the other hand, there is no corresponding superconformal group and the constraints on a SCKV mean that it differs from a non-conformal SKV only by a constant scale transformation.  In the following we shall go through each case in turn. It is straightforward to compute all the components of a SCKT starting from the leading even term. Except for $D=10$ it is simpler to use spinor notation and the Young tableaux calculus. This was introduced for $D=4$ in \cite{Howe:1981xy}
 where the tableaux were for the internal symmetry Lie algebra $(\gs)\gu (N)$. A single box with a dot (cross) then represents a covariant derivative $D_{\a i}\  (\bar D_{\adt}^i$). For the case in hand, however, it is more convenient to take the tableaux to represent spin representations. Again one can place either a cross or a dot inside a tableau to represent a spinorial derivative and then one can read off the representations of the internal symmetry algebra for a given component obtained from the top one by applying odd derivatives  in a simple fashion.

\subsection*{$D=3$}

In $D=3$ the spin group is $SL(2,\bbR)$ and the R-symmetry group is $O(N)$. The top component of a SCKT is a symmetric traceless tensor $K^{a_1\ldots a_n}$ obeying the constraint that applying an odd derivative $D_{\a i}$ produces a symmetric, traceless tensor-spinor which has $(n-1)$ vector indices and is also gamma-traceless. In spinor notation $K$ becomes a symmetric spinor with $2n$ indices, $K^{\a_1\ldots \a_{2n}}$, while its derivative has $(2n-1)$ symmetric spinor indices. The tableau for $K$ is
\be
K\sim \overbrace{\yng(8)}^{2n}   
\la{4.1}
\ee
and the constraint on $DK$ is 
\be
DK\sim \young(\cdot) \ \xz\  \overbrace{\yng(8)}^{2n}\ \sim \ \overbrace{\young(\hfil\hfil\hfil\hfil\hfil\hfil\hfil\hfil,\cdot)}^{2n}\ ,
\la{4.2}
\ee
in other words, the tableau with $(2n+1)$ symmetrised boxes in a row must vanish. No $SL(2,\bbR)$ tableau can have more than two rows, so any further $D$ must sit in the second row. Moreover, we are not interested in spacetime dependence at the moment, we only want to find the independent components of $K$ in a $\th$-expansion. So any pair of $Ds$ anti-commute, and since all of the spinor indices associated with the $Ds$ are in the same row, it follows that all the internal $\go(N)$ vector indices must be antisymmetrised. Thus, after $m$ steps we get
\be
D^mK \sim \overbrace{\young(\hfil\hfil\hfil\hfil\hfil\hfil\hfil\hfil,\cdot\cdot\cdot\cdot m)}^{2n}\ ,
\la{4.3}
\ee
or, in indices,
\be
D^m K\sim (D^mK)^{\a_1\ldots \a_{2n-m}}_{i_1\ldots i_m}\ ,
\la{4.4}
\ee
with symmetry on all of the spinor indices and antisymmetry on all of the internal indices.
Clearly $m\leq 2n$ and $m\leq N$. The spacetime constraint on $K$, namely that the symmetrised traceless part of $\del K$ vanishes, becomes
\be
\del ^{(\a_1\a_2} K^{\a_3\ldots \a_{2n+2})}=0\ .
\la{4.5}
\ee
This constraint then leads to similar constraints on the spacetime derivatives of all of the spinorial derivatives, 
\be
\del ^{(\a_1\a_2} (D^m K)_{i_1\dots i_m}^{\a_3\ldots \a_{2n-m+2})}=0\ .
\la{4.6}
\ee
This can also be represented by Young tableaux. If we apply a spacetime derivative to $K$ we get three tableaux, one with $(2n+2)$ boxes in the first row, one with $(2n+1)$ in the first row and one in the second, and one with $2n$ in the first two and $2$ in the second. It is $\del K$ in the representation corresponding to the first of these that must vanish, and this constraint then descends to all of the other components of $K$. Thus  $\del D^m K$ in the representation specified by  the tableau with $(2n+2)$ boxes on the first row and $m$ dotted boxes on the second must vanish.

As a simple example consider a SCKV \cite{Kuzenko:2010rp}. The components are $(K^{\a\b}, K^\a_i, K_{ij})$. The first component obeys the standard constraint for a conformal Killing vector in $D=3$, the second, $DK$, is a conformal Killing spinor $\k_i$, say,  \ie one satisfying
\be
\c_{\{a}\del_{b\}}\k_i=0\ ,
\la{4.7}
\ee
while the third, $(D^2 K)_{ij}$, is constant in $x$ and antisymmetric in $i j$. The leading component thus has spacetime conformal parameters, the second has $Q$ and $S$ supersymmetry parameters and the third is an $\gs\go(N)$ parameter.

\subsection*{$D=4$}
The situation in $D=4$ is roughly speaking the square of $D=3$. The spin group is $SL(2,\bbC)$ and  we use two-component spinors. A symmetric traceless SCKT $K^{a_1\ldots a_n}$ becomes an object with $n$ undotted and $n$ dotted spinor indices, symmetrised on both sets. It can be represented by a pair of tableaux,
\be
K\sim\left( \overbrace{\yng(8)}^{n}\ ,\ \overbrace{\yng(8)}^{n}\right) \ .
\la{4.7.1}
\ee
The constraints are
\begin{align}
DK&\sim\left( \overbrace{\young(\hfil\hfil\hfil\hfil\hfil\hfil\hfil\hfil,\cdot)}^{n}\ ,\ \overbrace{\yng(8)}^{n}\right) \nn\w1
\bar D K&\sim \left( \overbrace{\yng(8)}^{n}\ ,\ \overbrace{\young(\hfil\hfil\hfil\hfil\hfil\hfil\hfil\hfil,\xz)}^{n}\right)\ .
\la{4.7.2}
\end{align}
where the box with a dot corresponds to an undotted spinor and the box with a cross to a dotted one. $D$ also carries an internal $U(N)$ index in the fundamental $N$-dimensional representation, while $\bar D$ carries an anti-fundamental index. As in the $D=3$ case this means that the larger spin representations in $DK$ and $\bar D K$ are set to zero.
Acting on $K$ with $p$ $D$s will then give a tensor with $(n-p)$ symmetrised undotted indices and $p$ antisymmetrised $\gu(N)$ fundamental indices while leaving the dotted indices untouched. Similarly if we apply $q$ $\bar D$s we will get a tensor with $(n-q)$ dotted indices and $q$ antisymmetrised internal indices in the anti-fundamental representation of $\gu(N)$, while leaving the undotted indices untouched. Since we take $D$ and $\bar D$ to anti-commute for these purposes, it follows that the components $D^p\bar D^q K$ will have the form
\be
D^p\bar D^q K\sim (D^p\bar D^q K)^{\a_1\ldots \a_{n-p},\adt_1\ldots \adt_{n-q}, j_1\ldots j_q}_{i_1\ldots i_p}\ ,
\la{4.8}
\ee
or, in tableaux form,
\be
D^p\bar D^q K\sim \left(\overbrace{\young(\hfil\hfil\hfil\hfil\hfil\hfil\hfil\hfil,\cdot\cdot\cdot\cdot p)}^{n}\ ,\ 
\overbrace{\young(\hfil\hfil\hfil\hfil\hfil\hfil\hfil\hfil,\xz\xz\xz q)}^{n}\right)\ .
\la{4.8.1}
\ee
Clearly $p$ and $q$ must both be less or equal to the smaller of $n$ or $N$.  The reality of $K$ implies that $D^p\bar D^q K\sim \overline{(D^q\bar D^p K)}$. The spacetime constraint on $K$ is 
\be
\del^{(\a_1(\adt_1} K^{\a_2\ldots\a_{n+1})\adt_2\ldots \adt_{n+1})}=0\ .
\la{4.9}
\ee
Similar constraints hold for all the descendants,
\be
\del^{(\a_1(\adt_1} (D^p \bar D^q K)^{\a_2\ldots\a_{n-p+1})\adt_2\ldots \adt_{n-q+1})}=0\ ,
\la{4.10}
\ee 
where the internal indices have been suppressed.  

\subsection*{$D=6$}
 The situation in $D=6$ is a little more complicated. The spin group $Spin(1,5)$ is isomorphic to $SU^*(4)$, a non-compact form of $SU(4)$, for which  the Young tableaux are similar, while the R-symmetry group in the $N$-extended case is $Sp(N)$ (\ie $Sp(1)\cong SU(2)$ in the minimal ($N=1$) case). A vector may be written as an antisymmetric bi-spinor (note that pairs of anti-symmetrised spinor indices can be raised or lowered using the $\ve$ tensor), so that an $n$th rank symmetric traceless tensor corresponds to the $\gs\gu^*(4)$ Young tableau with $n$ columns and 2 rows,
\be
K\sim \ \overbrace{\yng(8,8)}^{n}\ .
\la{4.11}
\ee
The spinorial derivative $D_{\a i}$ corresponds to a single box with a dot in it to represent the $\gs\gp(N)$ index $i=1\ldots 2N$. Applying $D$ to $K$ we get two $\gs\gu^*(4)$ representations, but the larger one with the extra box in the first row vanishes by the SCKT constraint. This implies that when we act with $D^m$, again ignoring spacetime derivatives for the moment, the resulting Young tableaux will have extra dotted boxes siting below those of the original tableau, but of course there cannot be more than 4 rows overall. Thus we have
\be
D^m K\sim\ \sum \ \ \overbrace{\young(\hfil\hfil\hfil\hfil\hfil\hfil\hfil\hfil,\hfil\hfil\hfil\hfil\hfil\hfil\hfil\hfil,\cdot\cdot\cdot\cdot p,\cdot\cdot\cdot q)}^n\ ,
\la{4.12}
\ee
where $p+q=m, \ q\leq p\leq {\rm min}(n,2N)$, and the sum is over all possible tableaux compatible with these rules. These tableaux give the $\gs\gu^*(4)$ representations but they also give the internal symmetry ones, although not immediately as irreducible representations of $\gs\gp(N)$. Instead they are representations of $\gs\gu(2N)$ that can be further decomposed into irreducibles under $\gs\gp(N)$. The $\gs\gu(2N)$ tableaux are obtained by rotating the bottom two rows clockwise through ninety degrees and then reflecting about the vertical axis. Thus we get $\gs\gu(2N)$ diagrams of the form
\be
\young(\cdot\cdot,\cdot\cdot,\cdot\cdot,\cdot q,p)\ .
\la{4.13}
\ee
which can be decomposed into irreducible $\gs\gp(N)$ representations by removing the symplectic traces. The spacetime constraint on $K$ is $\del_{\{a_1} K_{a_2\ldots a_{n+1}\}}=0$. In terms of Young tableaux this means that the one in $\del K$ with $(n+1)$ columns and two rows must vanish. This constraint implies similar constraints for all of the descendants:  the component of  $\del D^m K$ in the representation corresponding to the tableaux for $\del D^m K$ obtained from that of $D^m K$ by appending one extra box in each of the first two rows must vanish.

\subsection*{{\sl Remarks on} $D=3,4,6$}

In the preceding subsections we have computed the components of a SCKT in a $\th$-expansion when the top component satisfies the standard constraint \eq{2.12}. One might ask whether this leads to an irreducible object or whether higher-order constraints (in fermionic derivatives) could be imposed. It is well-known that this is the case for a $D=4,N=4$ SCKV because the superconformal algebra is $\gp\gs\gl(4|4)$ rather than $\gs\gl(4|4)$. Consider the quantity $(D_{\a i}\bar D_{\adt}^i - \bar D_{\adt}^i D_{\a i})K^{\a\adt}:=k$; differentiating this with respect to $D$ we get
\be
D_{\a i} k\propto (N-4) \del_{\a\bdt} D_{\b i} K^{\b\bdt}\ .
\la{4.13.a}
\ee
On the right-hand side the object $\del_{\a\bdt} D_{\b i} K^{\b\bdt}$ is the $S$-supersymmetry parameter, so that it is possible to set $k=0$ only when $N=4$. In this case $k$ is actually the  parameter of the so-called $\gu(1)_Y$ algebra \cite{Intriligator:1999ff} which does not form a part of the superconformal algebra. Notice that this additional constraint is not implied by the standard SCKV constraint for $D=4,N=4$; it must be imposed separately.

It is possible that similar additional constraints could arise for higher-order SCKTs. In general we can arrange the $\th$-components of a general $D=4$ SCKT in a diamond-shaped array with each vertex labelled by a pair of integers, $(p,q); \ p,q\leq n$, connected by arrows representing the action of $\bar D$ or $D$ acting respectively to the left or right down the diagram, and ending at $(0,0)$. The vertex $(p,q)$ therefore represents a tensor with $p\  (q)$ symmetrised undotted (dotted) spinor indices, and $(n-p)\ ((n-q))$ antisymmetrised lower (upper) internal indices. For example, for $n=2$, we have the diagram:

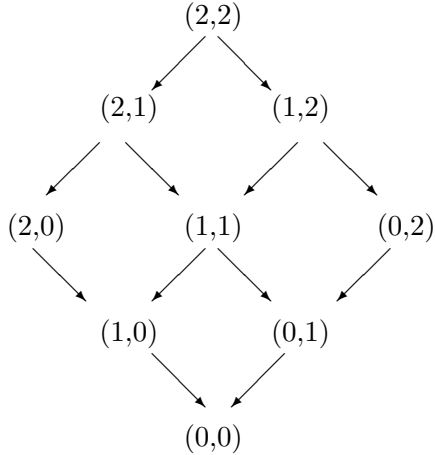
\be
\begin{picture}(400,200)
\put(202,180){(2,2)}
\put(210,175){\vector(-1,-1){20}}\put(215,175){\vector(1,-1){20}}
\put(170,145){(2,1)}\put(235,145){(1,2)}
\put(170,135){\vector(-1,-1){20}}\put(180,135){\vector(1,-1){20}}\put(245,135){\vector(-1,-1){20}}\put(255,135){\vector(1,-1){20}}
\put(135,100){(2,0)}\put(202,100){(1,1)}\put(275,100){(0,2)}\put(280,95){\vector(-1,-1){20}}
\put(145,95){\vector(1,-1){20}}\put(170,60){(1,0)}
\put(210,95){\vector(-1,-1){20}}
\put(215,95){\vector(1,-1){20}}\put(235,60){(0,1)}
\put(190,55){\vector(1,-1){20}}\put(240,55){\vector(-1,-1){20}}
\put(202,20){(0,0)}

\end{picture}
\la{4.13.b}
\ee
Consider the vertex $(1,1)$; it represents a tensor of the form $K_i^{\a \bdt j}$, so that its trace over the internal symmetry indices gives a vector $k^{\a\adt}$. Now this could be the leading component of a SCKV in the case that its supersymmetry variation does not contain any terms involving spacetime derivatives of components higher-up in the diagram. A simple computation shows that this can happen only for $N=6$. This result is related to the fact that some tensor representations of Lie superalgebras can be reducible but indecomposable even though the corresponding symmetry types are not in the non-super case \cite{Bars:1982se}. 
We shall come back to this point in section 6. 

The second comment we wish to make is that the definitions given above can clearly be generalised to the case where the leading components are spinorial; we might refer to such objects as superconformal Killing spinors (not to be confused with Killing spinors in supergravity). Thus for $D=3$ we could consider objects with an odd number of indices with similar tableaux to \eq{4.1}, and subject to similar constraints, while in $D=4$ one could have tensors with $(m,n), m\neq n$ (undotted,dotted) indices again subject to similar constraints to \eq{4.8}. In $D=6$ one could have objects with $n$ boxes in the first row and $(n-1)$ in the second again with the derivatives required to sit underneath the first two rows. It is not clear what the geometrical meaning of such objects is, but clearly they  
exist and should give rise to representations of  superconformal algebras.
\subsection*{$D=5$}

In five dimensions, superconformal symmetry only exists for $N=1$ \cite{Nahm:1977tg}. One can understand this from the fact that there is only one possible superconformal group, the exceptional Lie supergroup $F(4)$. It is easy to confirm this by a direct calculation. A putative SCKV $K^a$ satisfies the standard constraint\footnote{$D=5,N=1$ SCKVs were discussed previously in \cite{Kuzenko:2006mv}.}
\be
D_{\a i} K^a=i(\c^a)_{\a\b}\h_{ij} K^{\b j}=-i(\c^a)_{\a\b} K_i^{\b}\ ,\qquad \a,\b=1,\ldots 4: \qquad i,j=1\ldots 2N\ ,
\la{4.13.1}
\ee
where $\h_{ij}$ is the symplectic ``metric'' for the R-symmetry group $Sp(N)$. Differentiating this with respect to another odd derivative and taking the anti-commutator we find
\be
\del_a K^b=L_a{}^b + 2\d_a{}^b S;\qquad D_{\a i} K^{\b j}=\d_\a{}^\b L_i{}^j + \d_i{}^j ( L_\a{}^\b + \d_\a{}^\b S)\ ,
\la{4.13.2}
\ee
where $S$ is the scale parameter, $L_a{}^b$ the Lorentz parameter,  $L_\a{}^\b$ its spinorial counterpart, and where $L_{ij}=L_{ji}$ is the $\gs\gp(N)$ parameter. Differentiating the second of these equations and using the supersymmetry algebra again we find
\be
i\h_{ij}(\c^a)_{\a\b} \del_a K_k^\c=\left(\d_\a{}^\c(D_{\b j} L_{ik} + \h_{ik} D_{\b j} S) +\h_{jk}D_{\a } L_\b{}^\c\right) + (\a i\leftrightarrow \b j)\ .
\la{4.13.3}
\ee
There are three possible spinor representations in this equations, spinor, gamma-traceless vector-spinor and gamma-traceless tensor-spinor. It is not difficult to verify that the last two must be zero so we are left with just simple spinors. We set 
\begin{align}
D_{\a i}L_{jk}&=2\h_{i(j} \l_{\a k)} \nn\w1
D_{\a i} S&= \s_{\a i}\nn\w1
D_{\a i} L_{\b\c}&= 2\h_{\a(\b} \zeta_{\c) i}\ ,
\la{4.13.4}
\end{align}
while we can write the left-hand-side of \eq{4.13.3} as
\be
i\h_{ij}(\c^a)_{\a\b} \del_a K_{\c k}=(\c^a)_{\a\b} (\c_a)_\c{}^\d \r_{\d k}=-\h_{ij}\left(2\h_{\c(\a} \r_{\b) k}+\h_{\a\b} \r_{\c k}\right)\ ,
\la{4.13.5}
\ee
where $\h_{\a\b}$ denotes the symplectic ``metric'' on spinor space. The terms with $\h_{\a\b},\h_{\c\a}$ and $\h_{\b\c}$ must vanish separately, from which we find, from the terms with $\h_{\a\b}$,
\be
\h_{ij}\r_{\c k} + \h_{jk} \zeta_{\c i} + \h_{ki} \zeta_{\c j}=0\ .
\la{4.13.6}
\ee
If $N>1$ this equation has no non-trivial solution and one can use the $\h_{\b\c}$ equation to show that all of the spinors must vanish. On the other hand, if $N=1$, \eq{4.13.6} implies that $\r=\zeta$, and the $\h_{\b\c}$ equation then gives $\l=-\frac{3}{2}\r, \s=\half \r$. So in this case there is one spinor which can be identified with the S-supersymmetry parameter. For $N>1$, this analysis shows that $S$ is constant which in turn implies that there is no conformal boost. Hence we conclude that one can only have SCKVs for $N=1$ as anticipated.

For $N=1$ we can also have SCKTs obeying the usual constraint
\be
D_{\a i} K^{a_1\ldots a_n}=-in(\c^{\{a_1})_{\a\b} K_i^{a_2\ldots a_n\}}\ .
\la{4.13.7}
\ee
where $K^{a_1\ldots a_n}$ is traceless, and the tensor-spinor on the right is gamma-traceless. We can analyse the components of such an object by making use of the tableau calculus. For the case of $N=1$ we have
\be
K^a \sim \yng(1,1)\ ,
\la{4.13.8}
\ee
as a tableeau of the spin algebra $\gs\gp(2)$. In spinor indices $K^{\a\b}$ is symplectic-traceless, so that the tableau has the trace removed. Applying a derivative represented by a single-box tableau with a dot, we find
\be
DK\sim \young(\hfil,\hfil,\cdot) + \young(\hfil\cdot,\hfil)
\la{4.13.9}
\ee
The second diagram decomposes into the $4+ 16$ in $\gs\gp(2)$, and the constraint implies that the larger representation must be absent. We therefore have two spinor representations but in fact they are the same, as one can easily check. Because we can raise and lower spinor indices with the symplectic ``metric'' we only need one tableau, with a single box, as far as the spin group is concerned. We therefore write
\be
DK\sim \young(\hfil,\ast)
\la{4.13.10}
\ee
where the box with an asterisk does not carry an $\gs\gp(2)$ index, but does represent the $\gs\gp(1)$ internal symmetry doublet carried by $D_{\a i}$.  In other words, the derivative removes a box from the  original diagram \eq{4.13.8} and replaces it with an asterisked box which only represents the $\gs\gp(1)$ content. Applying a second $D$ we get
\be
D^2K\sim \young(\ast,\ast) \ ,
\la{4.13.11}
\ee
which represents a Lorentz scalar in the triplet representation of $\gs\gp(2)$, because the internal indices have the opposite symmetrisation properties to the spinors in the tableau.

This can be generalised very easily to SCKTs of arbitrary rank $n$. We have
\be
K\sim \overbrace{\yng(8,8)}^{n}\ ,
\ee
which implies
\be
DK\sim \overbrace{\young(\hfil\hfil\hfil\hfil\hfil\hfil\hfil\hfil,\hfil\hfil\hfil\hfil\hfil\hfil\hfil\ast)}^{n}\ ,
\ee
\be
D^2K\sim \overbrace{\young(\hfil\hfil\hfil\hfil\hfil\hfil\hfil\ast,\hfil\hfil\hfil\hfil\hfil\hfil\hfil\ast)}^{n}\  + \ \overbrace{\young(\hfil\hfil\hfil\hfil\hfil\hfil\hfil\hfil,\hfil\hfil\hfil\hfil\hfil\hfil\ast\ast)}^{n}\ ,
\ee
\be
D^3 K\sim
\overbrace{\young(\hfil\hfil\hfil\hfil\hfil\hfil\hfil\ast,\hfil\hfil\hfil\hfil\hfil\hfil\ast\ast)}^{n}\ ,
\ee
and, finally,
\be
D^4 K\sim
\overbrace{\young(\hfil\hfil\hfil\hfil\hfil\hfil\ast\ast,\hfil\hfil\hfil\hfil\hfil\hfil\ast\ast)}^{n}\ .
\ee
The diagram for $K$ is understood to be totally symplectic-traceless, and this property is inherited by the descendants. Since the internal indices on the same row are antisymmetrised there cannot be more than two of them, and hence the descent must stop at level four for $n>1$. Each component in the $\th$-expansion depends on $x$, and there are space-time constraints  that follow from the usual CKT constraint on $K$ itself. The expansion in $x$ for $D^m K$ will therefore go up to $x^{2n-m}$.
\subsection*{$D=10$}

In $N=1, D=10$ supersymmetry the spinors are Majorana-Weyl with sixteen components. We denote the gamma-matrices by $(\c^a)_{\a\b}$; they are symmetric on their spinor indices. Spinor indices cannot be raised or lowered so there is a corresponding set of gamma matrices with upper indices. It is no longer useful to think about the representations in a SCKT using Young tableaux, but instead we can work them out easily enough using Dynkin labels.\footnote{We use the term SCKT here even though there are no $S$-supersymmetry transformations or conformal boosts. This is because we still use the same basic constraint \eq{2.12}.} The basic representations have a 1 in the $k$th slot, where $k\in\{1,2,3,4,5\}$, with the other four labels being zero. They give $k$-forms for $k=1,2,3$ while the two 16-component spinor representations are given by $k=4,5$. If we take $\th^\a\sim(0,0,0,0,1)$ then $D_\a\sim (0,0,0,1,0)$. A symmetric, traceless $n$th rank tensor $K$ has Dynkin labels $(n,0,0,0,0)$. So $DK=(n,0,0,1,0)+((n-1),0,0,0,1)$. The constraint on an $n$th rank SCKT $K$ is
\be
D_\a K^{a_1\ldots a_n}= in(\c^{\{a_1})_{\a\b} K^{a_2\ldots a_n\}\b}\ ,
\la{4.14}
\ee
which in terms of Dynkin labels means that the larger $(n,0,0,1,0)$ representation is set to zero.
We can iterate this to obtain the $m$th descendant of $K$,
\be
D_{\a_1\ldots \a_m} K^{a_1\ldots a_n}\sim (\c^{\{a_1})_{[\a_1 \b_1}\ldots (\c^{a_m})_{\a_m]\b_m} K^{a_{m+1}\ldots a_n\}\b_1\ldots \b_m}\ .
\la{4.15}
\ee
Here $D_{\a_1\ldots \a_m}:=D_{[\a_1} D_{\a_2}\ldots D_{\a_n]}$, and the antisymmetrisation on the right-hand side is over the $\a$-indices only. It implies that the $\b$-indices must also be antisymmetrised and hence $K^{a_1\ldots a_{n-m}\b_1\ldots \b_m}$ is symmetric traceless on its even indices and antisymmetric on its odd indices. The descendants will be non-zero provided that $m\leq\ {\rm min}(n,16)$. The representations at each level need not be reducible but can be computed in terms of Dynkin labels fairly easily.

We give an example of such a SCKT for $n=6$. There are seven levels. The representations are, starting from the top: (60000); (50001); (40100); (31010); (30020)+(22000); (21010); (20100). So there are two representations at level $D^4 K$ but otherwise only one. This seems to be the case for all $n$ although we do not have a complete proof.

For an SCKV the derivative constraint on $K^a$ is the standard conformal one, but if one differentiates this with $D_\a$ and uses the defining constraint one easily finds that both the scale and Lorentz parameters are constant and that $K^\a$ is also constant. This confirms that there are no conformal boosts or $S$-supersymmetry transformations in $K^a$. For the higher rank SCKTs the generalised scale and Lorentz parameters are no longer constant but obey stronger constraints than in the case where there are conformal boosts and $S$-supersymmetry. For example, for a second-rank tensor, the expansion of the leading component goes up to $x^2$ whereas for the true conformal case it would extend to $x^4$. Thus the only difference between these objects and SKTs is that the Lorentz group is extended by a scale transformation.

\section{Analytic superspace}
In this section we shall discuss SCKTs in $D=3,4$ and 6 in the context of analytic superspaces. These are superspaces with fewer odd coordinates than the associated conventional (Minkowski) superspaces, such that superfields on these spaces correspond to fields on Minkowski superspace satisfying constraints with respect to the odd derivatives. These so-called Grassmann- (or G-) analytic superfields generalise the notion of chiral superfields. Typically they also depend on additional internal even coordinates. They were first introduced in the physics literature as harmonic \cite{Rosly,Galperin:1984av,Galperin:1984bu} or projective superspaces \cite{Karlhede:1984vr,Lindstrom:2008gs} in four dimensions. More general treatments were later developed in \cite{Lukierski:1988vw,Hartwell:1994rp,Howe:1995md} where it was found convenient
to work in complexified superspaces defined as cosets of the superconformal groups with parabolic isotropy groups: these are flag supermanifolds \cite{Manin:1988ds,Harnad:1987xq,Harnad:1995zy}.  All the fields are taken to be holomorphic, and we shall usually work on some open subset in the spacetime sector as the cosets themselves are compact (in the even directions). The spaces we shall consider all contain standard complexified Minkowski space as a component of the purely even part, and indeed there is a formal resemblance to Minkowski spaces considered as cosets of the conformal groups.\footnote{One can consider these spaces as supersymmetric versions of twistor geometry, see, for example, \cite{Penrose:1986ca,Baston:1989vh,Ward:1990vs}.} They have additional even sectors, cosets of the R-symmetry groups, and reduced number of odd coordinates compared to Minkowski superspace. The analytic superspace formalism we shall use is one in which local coordinates are employed for all of the coordinates including the internal and odd ones. We shall be interested in those for which the reduction in the number of odd coordinates is maximal, and we shall also restrict our attention to the simpler cases of $N$ even, for $D=3,4$. For examples of this formalism applied to $N=4$ superconformal field theory see, for example, \cite{Heslop:2003xu}.

$D=3$ harmonic superspace was introduced in \cite{Zupnik:1988wa} and later developed in \cite{Howe:1994ms}, while $D=6$ was first discussed as projective superspace in \cite{Grundberg:1984xr} and as harmonic superspace in \cite{Howe:1985ar}. A detailed study of $D=6$ superconformal field theory in an analytic superspace setting can be found in
\cite{Heslop:2004du}.

Let us recall that $D=4$ complex Minkowski space can be regarded as the Grassmannian of two-planes in $\bbC^4$. It is a quotient of the complexified conformal group by the isotropy group that preserves a two-plane; this is the group of $4 \xz 4$ matrices that consists of $2\xz 2$ blocks with a zero block in the top right-hand corner. The map between Minkowski space $M$ and the conformal group can be represented by
\be
M\ni x \rightarrow \left( \barr{cc} 1 & x\\ 0 & 1 \earr\right)\ ,
\la{5.1}
\ee
where $x$ is the $2\xz 2$ matrix $\ x^{\a\a'}$ in two-component spinor notation. A conformal Killing vector $K^{\a\a'}$ satisfies
\be
\del_{\a\a'} K^{\b\b'} = a_1(\d_{\a}{}^{\b} \del_{\c\a'} K^{\c\b'}+\d_{\a'}{}^{\b'} \del_{\a \c'} K^{\b\c'})+b_1\, \d_\a{}^\b \d_{\a'}{}^{\b'} \del\cdot K\ ,
\la{5.1.1}
\ee
where $a_1,b_1$ are constants that can be computed by consistency (see below). This 
can be generalised to a CKT $K^{\a_1\ldots \a_n,\a'_1\ldots\a'_n}$ as follows,
\begin{align}
\del_{\a\a'} K^{\b_1\ldots \b_n,\b'_1\ldots\b'_n}& =a_n( \d_\a^{(\b_1} \del_{\c\a'} K^{\b_2\ldots \b_n)\c,\b'_1\ldots\b'_n} + 
\d_{\a'}^{(\b'_1} \del_{\a\c'} K^{\b_1\b_2\ldots \b_n,\b'_2\ldots\b'_n)\c'}) \nn\w1
&+b_n\,\d_\a{}^{(\b_1} \d_{\a'}{}^{(\b'_1}\del_{\c\c'} K^{\b_2\ldots\b_n)\c,\b'_2\ldots\b'_n)\c'}
\la{5.2}
\end{align}
where the parentheses apply separately to the $\b$ and $\b'$ indices.

One has similar constructions for $D=3$ where $x$ is a $2\xz 2$ symmetric matrix $x^{\a\b}$ and in $D=6$ where $x$ is a $4\xz 4$ antisymmetric matrix. 

The above constructions generalise rather easily to the super case.

\subsection*{$D=4$}

For even $N=2M$ in $D=4$, analytic superspace is the super Grassmannian of $(2|M)$-planes in $\bbC^{4|2M}$, complex flat superspace with 4 even and $2M$ odd directions. It is actually more convenient to change the ordering of even and odd to $\bbC^{2|M|2|M}$, \ie 2 even, $M$ odd, 2 even and $M$ odd. Analytic superspace is the coset space of the complex superconformal group $SL(4|2M)$ with isotropy subgroup consisting of matrices of the form
\be
\left(\ba{cc} L & 0\\ L&L\ea\right)\ ,
\la{5.2.1}
\ee
where each $L$ denotes a $(2|M) \xz (2|M)$ matrix, and such that the full matrix is an element of the superconformal group. A point in analytic superspace can then be represented by an element of the group of the form
\be
M\ni X \rightarrow \left( \barr{cc} 1 & X\\ 0 & 1 \earr\right)\ ,
\la{5.2.2}
\ee
so $x^{\a\a'}$ in \eq{5.1} is replaced by $X^{A A'}$. Here $A=(\a,a),\ A'=(\a',a')$ are super-indices with $\a,\a'=1,2$ while $A,A'$ run form 1 to $M$.\footnote{In section 5 only we use $\a'$ instead of $\adt$ for $D=4$ dotted spinor indices, while $a,\,a'$ are internal indices. We take $\a,\a'$ to be even indices while $a,a'$ are odd.} The super-coordinates $X^{AA'}=(x^{\a\a'},\x^{\a a'},\x^{a \a'},y^{a a'})$, where $x^{\a\a'}$ are the even spacetime coordinates, $\x^{\a a}$ and $\x^{a\a'}$ are odd coordinates and $y^{a a'}$ are internal coordinates representing (a patch of) the internal Grassmannian of $M$-planes in $\bbC^{2M}$.  Since analytic superspace is defined as a coset, we can straightforwardly work out the effect of  infinitesimal superconformal transformation on the local coordinates; it is given by
\be
\d X^{AA'}=b^{AA'} + a^A{}_B X^{BA'} + X^{A B'} d_{B'}{}^{A'}+ X^{AB'} c_{B'B} X^{BA'}\ ,
\la{5.2.3}
\ee
corresponding to the element
\be
z=\left( \barr{cc} -a & b\\ -c& d \earr\right)
\la{5.2.4}
\ee
of the Lie superalgebra which we could take to be $\gg\gl(4|2M)$.  For $M\neq 2$, \ie   $N\neq 4$, one can take out the supertraces from both $a$ and $d$ leading to the observation that $\str(z)$ does not act on the coordinates. So in these cases we can take the superalgebra to be $\gs\gl(4|2M)$. For $N=4$, however, $\str(z)$ does act and defines the $\gu(1)_Y$ parameter. We shall therefore have to impose $\str(z)=0$ as an additional constraint in this case. We shall discuss this further below in the context of reducibility problems.  The right-hand side of \eq{5.2.3} can be identified as a SCKV $K^{AA'}$ and satisfies the differential equation
\be
\del_{AA'} K^{BB'} =a_1( \d_{A}{}^{B} \del_{CA'} K^{C B'}+\d_{A'}{}^{B'} \del_{A C'} K^{B C'})+ b_1\,\d_A{}^B\d_{A'}{}^{B'} \del_{CC'} K^{C C'}\ .
\la{5.2.5}
\ee
As in the bosonic case this can be generalised to higher-rank tensors. A SCKT  is a tensor on $N$-extended $D=4$ analytic superspace of the form $K^{A_1\ldots A_n,A'_1\ldots A'_n}$, graded symmetric on primed and unprimed indices, obeying the constraint
\begin{align}
\del_{AA'} K^{B_1\ldots B_n,B'_1\ldots B'_n}&= a_n( \d_A{}^{(B_1}\del_{CA'} K^{B_2\ldots B_n)C,B'_1\ldots B'_n} +
\d_{A'}{}^{(B'_1}\del_{AC'} K^{B_1\ldots B_n,B'_2\ldots B'_n)C'})\nn\w1 
&+b_n\, \d_A{}^{(B_1}\d_{A'}{}^{(B'_1}\del_{CC'} K^{B_2\ldots B_n) C,B'_2\ldots B'_n)C'}\ ,
\label{5.2.6}
\end{align}
where the bracket refer to the sets of $B$ and $B'$ indices. As well as the constants $a,b$ that appear in the above equations there are also sign factors which are necessary to maintain covariance.\footnote{Throughout the paper all tensorial equations are understood to be covariant and we do not include Grassmann sign factors explicitly; it is always possible to do this.} The constants are given by
\be
a_n=\frac{1}{t_n} \qquad b_n=-\frac{1}{(t_n)^2}\ ,
\la{5.3.01}
\ee
where 
\be
t_n=\frac{n-1+t}{n}\ ,
\la{5.3.a}
\ee 
$t$ being the super-trace over the primed or unprimed indices, \ie $t=2-M$. (These are also valid in the bosonic case when $M=0$.) It is evident from these formulae that the coefficients are singular whenever $t_n=0$. This is not a real problem because the partial divergences themselves contain factors involving the super-traces, so that the zeroes cancel out. An example is given by $N=4, n=1$. In this case a super-conformal Killing vector is given by the right-hand side of \eq{5.2.3}. Differentiating with respect to $X$ is then consistent with \eq{5.2.5} because the partial divergences in the $a_1$ term give factors of $t$ while the full divergence in the $b_1$ term gives a factor of $t^2$. Similar remarks apply in the $D=3$ and $6$ cases.

It is rather easy to solve the SCKT equation \eq{5.2.6}. The solution can be represented by a diamond structure of the type \eq{4.13.b}, but where now the $k$th row corresponds to the terms with $X^k$, while the vertex $(p,q)$ indicates (for an $n$th rank SCKT) that the parameter has $p(q)$ free, symmetrised contravariant unprimed (primed) indices and, correspondingly, $(n-p)$ ($(n-q)$) covariant indices that are contracted with the $X$s. As an example, consider $n=2$:
\begin{align}
K^{AB,A'B'}&=b^{AB,A'B'} + \left(a^{AB}{}_C{}^{A'} X^{C B'}+X^{AC'} d^B{}_{C'}{}^{A'B'}\right) \nn\w1
&+ \left(a^{AB}{}_{CD} X^{CA'} X^{D B'} +  X^{AC'} b_{C'}{}^{B A'}{}_{C} X^{C B'} + X^{AC'} X^{BD'} d_{C'D'}{}^{A'B'}\right)\nn\w1
&+ \left(a^A{}_{C D D'}X^{B D'} X^{CA'}X^{D B'} + X^{AC'} X^{B D'} X^{D A'} d_{D C'D'}{}^{B'}\right)\nn\w1
&+X^{A C'} X^{BD'} b_{C'D'CD} X^{CA'}X^{DB'}\ ,
\label{5.3.b}
\end{align}
where the indices $AB$ and $A'B'$ are understood to be symmetrised. For each power of $X$ except the zeroth and fourth, there are redundant parameters. For example, in the $X^1$ terms, there can be traces in both $a$ and $d$ only one of which is independent.

\subsection*{$D=3$}

In $D=3$  the spacetime conformal group is symplectic ($Sp(2)$), the R-symmetry group is $O(N)$  and the superconformal groups are orthosymplectic, $SpO(2|N)$ (as mentioned above, this is the same as $OSp(N|2)$ but we have written it in the reverse order to indicate that $Sp(2)$ is the spacetime part). When $N=2M$ we can define (complex) analytic superspace as follows: it is the space of isotropic $(2|M)$-planes in $\bbC^{4|2M}$. By isotropic we mean that any pair of vectors belonging to  such a plane have vanishing scalar product with respect to the super-symplectic tensor $\cJ$ which we can take to have the form
\be
\cJ=\left(\ba{cc} 0&I\\J&0\ea\right)\ ,
\la{5.3.1}
\ee 
where we have split the full space $\bbC^{4|2M}$ into two halves each with dimension $(2|M)$, and where
\begin{align}
I&=\left(\ba{cc} 1_2&0\\0&1_M\ea\right)\nn\w1
J&=\left(\ba{cc} -1_2&0\\0&1_M\ea\right)\ 
\label{5.3.2}
\end{align}
where the non-zero entries are the identity matrices with the  indicated dimensions. Alternatively, we can say that the $(4|M)\xz (4|M)$-matrix with only non-zero element $X$ in the top right (as in \eq{5.2.2}) belongs to the super Lie algebra $\gs\gp\go(2|2M)$. This consists of matrices $L$ of the form \eq{5.2.4} satisfying
\be
L \cJ + \cJ L^{st}=0\ ,
\la{5.3.3}
\ee
where the super-transpose is the matrix transpose with an additional minus sign when the first index is odd and the second even.
This then implies that the  local coordinates for analytic superspace, for $N=2M$, are $X^{AB}=X^{BA}$, where the symmetry is graded. We have $X^{AB}=(x^{\a\b},\x^{\a b},y^{ab})$, where $x^{\a\b}$ are the complex spacetime coordinates, $\x^{\a b}$ the odd co-ordinates and $y^{ab}$ the internal coordinates, with the internal indices running from 1 to $M$..  A SCKT $K^{A_1\ldots A_{2N}}$ is a totally graded-symmetric $2n$th rank tensor satisfying
\be
\del_{A_1 A_2} K^{B_1\ldots B_{2n}}= a_n\, \d_{(A_1}{}^{(B_1} \del_{A_2) C} K^{B_2\ldots B_{2n} )C}+b_n\,
\d_{A_1}{}^{(B_1} \d_{A_2}{}^{B_2}\del_{C D}K^{B_3\ldots B_{2n})CD}\ ,
\la{5.4}
\ee
where the brackets denote graded symmetrisation. The constants $a,b$ are given by
\be
a_n=\frac{4n}{t+2n}\qquad b_n=-\frac{2n(2n-1)}{(t+2n)(t+2n-1)}\ ,
\la{5.4.01}
\ee
where $t=2-M$ is again the supertrace.

The solution to equation \eq{5.4}  is
\be
K^{A_1\ldots A_{2n}}=\sum_{m=0}^{m=2n} X^{A_1B_1}\ldots X^{A_m B_m} a_{B_1\ldots B_m}{}^{A_{m+1}\ldots A_{2n}}\ ,
\la{5.4.02}
\ee
where it is understood that the $A$-indices are totally graded-symmetrised. Note that in this case, there are no redundant parameters so that it is not necessary to subtract out supertraces. Thus, although there could be values of $n$ and $N=2M$ for which there are indecomposable representations of $\gg\gl(2|M)$ this does not cause any difficulties for SCKTs in $D=3$.
\subsection*{$D=6$}

$D=6$ is similar in some ways to $D=3$ with the  difference that the roles of the symplectic and orthogonal groups are interchanged. The superconformal groups are $OSp(8|N)$, where the symplectic groups are now the R-symmetry groups. Analytic superspace is the space of isotropic $(4|N)$ planes in $\bbC^{8|2N}$, but now the internal and spacetime sectors are interchanged compared with the $D=3$ case. The ortho-symplectic metric $\cG$ is 
\be
\cG=\left(\ba{cc} 0&I\\J&0\ea\right)\ ,
\la{5.3.001}
\ee 
where we have split the full space $\bbC^{8|2N}$ into two halves each with dimension $(4|N)$, and where
\begin{align}
I&=\left(\ba{cc} 1_4&0\\0&1_N\ea\right)\nn\w1
J&=\left(\ba{cc} 1_4&0\\0&-1_N\ea\right)\ 
\label{5.3.002}
\end{align}
where the non-zero entries are the identity matrices with the  indicated dimensions. The super Lie algebra $\go\gs\gp(8|N)$ consists of matrices $L$ of the form \eq{5.2.4} satisfying
\be
L \cG+ \cG L^{st}=0\ ,
\la{5.3.003}
\ee
and the coordinate matrix $X$ is an element of this super-algebra with non-zero upper right elements only. This is similar to the $D=3$ case but with the internal and spacetime even dimensions interchanged. Thus, in this case,
the local coordinates $X^{AB}=(x^{\a\b},\x^{\a b},y^{ab})$, $\a=1,\dots 4,\ a=1,\ldots N$, are graded antisymmetric, so that we indeed have six even spacetime coordinates. 

A SCKT now has the form $K^{A_1 A_2,B_1 B_2,C_1 C_2,\ldots}$, with graded antisymmetry on each pair and symmetry under the interchange of pairs. Moreover, $K$ vanishes if it is graded-antisymmetrised  on any three indices (\eg $K^{[A_1 A_2,B_1] B_2,C_1 C_2,\ldots}=0$). In other words the symmetry structure corresponds to the tableaux \eq{4.11}, but where the symmetrisations are understood to be graded.  A SCKT $K$ satisfies the constraint
\begin{align}
\del_{A_1 A_2} K^{B_1 B_2,C_1, C_2,\ldots} &=(a_n\, \d_{[A_1}{}^{[B_1} \del_{A_2] D} K^{B_2] D,C_1 C_2,\ldots} +
(n-1)\  {\rm terms})\nn\w1
&+ b_n\,(\d_{[A_1}{}^{[B_1}\d_{A_2]}{}^{B_2]}\del\cdot K^{C_1C_2,\ldots}+ {\rm cyclic})\nn\w1
&-\frac{6b_n}{n+1}(\sum \d_{[A_1}{}^{[B_1}\d_{A_2]}{}^{B_2} \del\cdot K^{C_1 C_2],D_1D_2,\ldots})\ ,
\la{5.5}
\end{align}
where in the second line the cyclic sum is over the $n$ pairs, and where the sum in the third line is over all distinct pairs of pairs, \ie $\half n(n-1)$ terms altogether. In the expression on the third line for each selected pair of pairs there is total graded antisymmetrisation. It can be checked that these terms are necessary to ensure that the (graded) symmetry structure of the tableau \eq{4.11} holds for the $b$ terms, while the $a$ terms take care of themselves. We have used the dot notation to denote the divergence with respect to a given pair of indices. The coefficients are given by
\begin{align}
a_n&=\frac{4}{t+n-3}\nn\w1
b_n&=\frac{-(n+1)}{(t+n-2)(t+n-3)}\ .
\la{5.6}
\end{align}

\subsection*{\sl Reducibility problems in ASS}

In $D=4$ Minkowski superspace we saw that there can be cases where the standard SCKT constraint \eq{2.12} does not lead to an irreducible system and that further constraints can be imposed. Here we briefly discuss this problem in ASS. In all cases the Lie superalgebra that acts on the ASS indices ($A,A'$) is $\gg\gl(P|Q)$. We can study reducibility problems for finite-dimensional tensor representations with $r$ contravariant and $s$ covariant indices, taken to be totally symmetric on both sets, and totally super-traceless, by looking for tensors of this type which have $p$ factors of the unit matrix, $p\leq {\rm min} (r,s)$, but which remain super-traceless. It is straightforward to derive the following formula for when this can happen:
\be
r+s-p=1-t
\la{5.7}
\ee
where $t$ is the relevant super-trace, $t=P-Q$. When this equation is satisfied one can in principle have indecomposability problems.

Let us consider first the case $D=4$. Here $t=2-M$, $M=N/2$, because both the primed and unprimed indices are acted on by $\gg\gl(2|M)$ Lie superalgebras. The formula becomes
\be
N=2((r+s)+1-p)
\la{5.8}
\ee
For $n=1$, \ie SCKVs, we must also have $r=s=p=1$ to get a solution to \eq{5.8}, and this implies $N=4$. This stems from the fact that the matrix parameters in \eq{5.2.3} cannot be made super-traceless for $M=2$. However, we can impose the constraint that ${\rm str}(z)=0$, where $z$ is given by \eq{5.2.4}, so that the full superalgebra is $\gs\gl(4|4)$. This additional constraint is equivalent to the extra condition that was imposed in super Minkowski space to get rid of the $\gu(1)_Y$ transformation. Once this has been done, the unit matrix in $z$, which has vanishing super-trace, does not act on $X$, and this implies that the algebra is $\gp\gs\gl(4|4)$. For all other values of $N$ the super-traces can be removed from $a$ and $d$, so that there are two ``scale'' transformations of $X$ only one of which is independent. Since the unit matrix in $z$ does not act, we can simply set it equal to zero, so that the algebra is $\gs\gl(4|N)$.

Now consider $n=2$. In equation \eq{5.3.b}, the $a$ and $d$ parameters have $(r,s)=(2,1),(2,2),(1,2)$ with respect to the unprimed (primed) indices respectively. For the $(1,2)$ and $(2,1)$ cases we have $p=1$ so that there is a problem for $N=6$, while for $(r,s)=(2,2)$ we can choose $p=2$ and still have $N=6$. These reducibility problems correspond to the $n=2$ case discussed previously in super Minkowski space where we found a problem of this sort precisely for $N=6$. The problem can be resolved in a similar fashion to the $N=4$ case we have just discussed. We can impose further constraints to get rid of half of the super-traces in the $a$ and $d$ parameters, but we are then left with a residual invariance similar to the projective symmetry in $N=4$.

In the following section we shall address the issue of indecomposable representations for SCKTs in super-twistor spaces where the superconformal algebras act linearly.

\section{Components of superconformal Killing tensors}

In the purely even case an $n$th rank conformal Killing tensor on flat spacetime satisfies
\be
\del_{\{a_1} K_{a_2\ldots a_{n+1}\}}=0\ .
\ee
This equation can be solved as a finite power series in $x$ with constant coefficients which can be assembled into a representation of the conformal group, $O(2,D)$. For example, when $n=1$, the components of a CKV together form the adjoint representation of $\go(2,D)$. In the general case one can use the representation of $D$-dimensional spacetime as a surface in flat $(D+2)$-dimensional space with two timelike directions. One can then explicitly show that an $n$th rank CKT has components that fall into the representation of the conformal group given by the Young tableau \cite{Eastwood:2002su}
\be
\overbrace{\yng(7,7)}^{n}\ ,
\la{6.2}
\ee
where the representation is taken to be irreducible, \ie all traces are removed. Thus the original (irreducible) one-row $n$-box tableau that represents the CKT in spacetime determines the two-row tableau \eq{6.2} as a representation of the conformal group.

This picture does not generalise to the super case because one does not have an analogous superconformal embedding of super Minkowski space. This is not surprising given that there are only superconformal groups in $D=3,4\,\&\, 6$ (apart from the exceptional $N=1, D=5$ case). However, the components of a SCKT can still be assembled into a representation of the appropriate superconformal algebra. If we complexify the (super)spaces involved, we can represent the components of a SCKT as a tensor on the appropriate super-twistor spaces which are $\bbC^{4|N}$ for $D=3,4$ and $\bbC^{8|2N}$ for $D=6$. These are the fundamental representation spaces for the supergroups $SpO(2|N),\ SL(4|N)$ and $OSp(8|N)$ respectively.\footnote{Super-twistors were introduced in \cite{Ferber:1977qx} and have been used in the discussion of  superparticles in a manifestly superconformal context \cite{Bengtsson:1987ap,Townsend:1991sj}.}

\subsection{$D=3$}

The simplest case is $D=3$. Let us start with the purely even case again. An $n$th rank CKT is given by the one-row $2n$-box tableau in $\gs\gl(2)$:
\be
K\sim \overbrace{\yng(8)}^{2n} \ ,  
\la{6.a}
\ee
satisfying the constraint \eq{4.5}. Solving this we find that the components of $K$ can be represented by the same diagram, but this time  for the conformal algebra $\gs\gp(2)$.  We know that $\gs\gp(2)\cong\go(5)$ and it is easy to check that this representation in $\gs\gp(2)$ is indeed the same as the irreducible two-row Young tableau in $\go(5)$.

Now consider the supersymmetric $D=3$ case for arbitrary $N$. Again the SCKT is given by \eq{4.1} as a diagram in $\gs\gl(2)$, and we know the supersymmetric descendants from the discussion given in section 4. All of these satisfy conformal spacetime constraints and can be computed as representations of $\gs\gp(2)$; in fact the $m$th descendant will correspond to the representation of $\gs\gp(2)$ with $(2n-m)$ boxes. In addition this descendant will transform as an $m$th rank totally antisymmetric tensor under $\go(N)$. Putting all this together, we can see that all of the components of the SCKT determined by $K$ can be assembled into a tensor of $\gs\gp\go(2|N)$ given by exactly the same tableau but now considered as a super Young tableau for the $D=3$ $N$-extended superconformal algebra. Super Young tableaux \cite{Bars:1982se} are interpreted in a similar way to non-super ones except for the fact that (anti-)symmetrisation is replaced by super-(anti-)symmetrisation. For example, a tableau with one row and $n$ columns represents a graded symmetric tensor; when expanded out into even and odd components, the even indices are symmetrised while the odd ones are antisymmetrised.

In the $D=3$ case no reducibility problems arise; the super-tensors defined by \eq{4.1} are all irreducible.

It is straightforward to relate this discussion to the ASS one. We can split a super-twistor index for $\bbC^{(4|N)}$ into a pair of $\bbC^{(2|M)}$ indices (when $N=2M$) as follows:
\be
Z^{\cA}=(Z_{A}, Z^{A})\  ,
\la{4.1.1}
\ee
then
\be
K^{\cA_1\ldots \cA_{2n}}\rightarrow \left(K_{A_1A_2\ldots A_{2n}},\ldots ,K_{A_1\ldots A_m}{}^{A_{m+1}\ldots A_{2n}},\ldots ,K^{A_1\ldots A_{2m}}\right)\ .
\la{4.1.2}
\ee
Clearly there is a one-to-one correspondence between these components of $K$ and the $a$-parameters in \eq{5.5}.

\subsection{$D=4$}

The super-conformal algebras for $D=4$ are $\gs\gl(4|N)$ (except for the special case $N=4$ where it is $\gp\gs\gl(4|4)$). Because there is no straightforward generalisation of the epsilon tensor in the super case a single type of tableau box does not suffice to describe all representations \cite{Bars:1982se}. Instead it is necessary to introduce two basic single-box tableaux to denote fundamental and anti-fundamental representations, corresponding to contravariant and covariant indices for the vector space $\bbC^{4|N}$. (So for $N=0$ the latter would be a one-column three-column tableau). We distinguish a covariant box by placing a bullet in it, $\young(\bullet)$, while a general tableau will have left and right sections with former corresponding to covariant indices and the latter to contravariant ones. In order to describe irreducible representations it is necessary to impose the requirement that all super-traces between contra- and co-variant indices are removed. An element of the $\gs\gl(4|N)$ super-algebra is given by a super-traceless tensor in $\bbC^{4|N}$ with one upper and one lower index and so has the tableau 
\be
 \young(\bullet)\yng(1)\ .
\la{6.0.1} 
\ee
This can be immediately generalised to an $n$th rank SKCT: such an object is given by a tableau of the form
\be
\overbrace{\young(\bullet\bullet\bullet\bullet\bullet\bullet)}^{n}\!\overbrace{\yng(6)}^{n}
\la{6.0.2}
\ee
A superconformal Killing vector, for the case $n=1$, has one index of each type and can be written $K^{\cA'\cB}$, or $K_{\cA}{}^{\cB}$, where an upper $\cA$ is a contravariant $\bbC^{4|N}$ super-index, and $\cA'$ a covariant one which can also be written as a lower unprimed index. The super-trace of $K_{\cA}{}^{\cB}$ is taken to vanish. An $n$th-rank SCKT is a tensor of the form
\be
K^{\cA'_1\ldots \cA'_n,\cB_1\ldots\cB_n}\cong K_{\cA_1\ldots \cA_n}{}^{\cB_1\ldots\cB_n}\ ,
\la{6.0.3}
\ee
totally graded-symmetric on each set of indices and totally super-traceless.

The above discussion is not complete because there can be reducibility problems \cite{Bars:1982se}. This means that in some cases there are super-traceless tensors that are reducible but indecomposable because of the existence of sub-representations that cannot be removed in a manifestly covariant way. The simplest example is $N=4, n=1$. In this case the unit tensor, $\d_\cA{}^\cB$, has vanishing super-trace but cannot be subtracted from $K_\cA{}^\cB$. This example just corresponds to the $N=4$ Lie superalgebra, so that the unit tensor has to be modded out to obtain $\gp\gs\gl(4|4)$.

Reducibility problems can occur for higher values of $n$ according to the formula
\be
N=2n+3-p\ ,
\la{6.0.4}
\ee
where $p\leq n$ denotes the number of unit tensor factors.\footnote{This is a special case of \eq{5.7} for $r=s=n, t=4-N$.}  In more detail, this means that one can have $n$th-rank tensors of the type of \eq{6.0.3} which contain $p$ factors of the unit tensor but which remain super-traceless. From \eq{6.0.4} it can be seen that problems of this type only arise for $N\geq 4$ and that for the $N=4$ case the only problem is the one just discussed with $n=1$. In particular, this implies that higher-order super-traceless tensors in $N=4$ are projectively invariant and so correspond to representations of $\gp\gs\gl(4|4)$.  Note, however, that when one attempts to decompose tensors of the above type that are not super-traceless into their irreducible super-traceless parts one can run into these problems more than once. For example, for $N=4,n=2$ one can extract the traceless part leaving the unit tensor times an $n=1$ object which is itself indecomposable, although this turns out not to be a problem in the algebraic context discussed in 6.4.

In the $D=4$ case it is also possible to make contact with the ASS discussion. We set
\be
Z^{\cA}=(Z_{A}, Z^{A'})\ \ \ ,\ \ \  Z_{\cA}=(Z^{A}, Z_{A'})\ ,
\la{6.0.5}
\ee
which allows us to decompose any tensor in terms of $\gg\gl(2|M)\oplus \gg\gl(2|M)$ representations. For example, for $n=2$, the components of $K_{\cA\cB}{}^{\cC\cD}$ are 
\begin{align}
& K^{AB C'D'}\nn\w1
K^{AB}{}_{C}^{\ \ D'}\phantom{KKk}&\qquad\qquad\  K^A{}_{B'}{}^{C'D'}\nn\w1
K^{AB}{}_{CD}\qquad K^{A'}{}_{B'}{}^{C'}{}_{D}&\qquad  K^{A'}{}^{B}{}_{C}{}^{D'}\qquad K_{A' B'}{}^{C'D'}\nn\w1
K_{A'}{}^B{}_{CD}\phantom{KKk}&\qquad\qquad\  K_{A'B'}{}^{C'}{}_{D}\nn\w1
& K_{A'B'CD}\ .
\la{6.06}
\end{align}

These components can be matched to the $a,b$ and $d$ parameters in \eq{5.3.b}, with those in the $k$th row corresponding to the terms with $X^k$. For example, at $X^1$, we have $K^{AB}{}_C{}^{D'}\sim -a^{AB}{}_C{}^{D'}$ and $K^A{}_{B'}{}^{C'D'}\sim d^A{}_{B'}{}^{C'D'}$ (the minus sign orginates from the conventions for $z$ in \eq{5.2.4}). If we can remove the super-traces from over the central indices from $a,d$ then we can see that the combination
\be
({\rm str}(a)-{\rm str}(d))^{AD'}\ 
\la{6.07}
\ee
does not contribute to the ASS SCKT $K^{AB,A'B'}$ in \eq{5.3.b} and so can be set to zero. This corresponds to setting $({\rm str} K)^{AD'}=0$ in super-twistor space. If it is not possible to remove the super-traces then one can impose the this super-traceless condition on $K$, but then one will still be free to adjust $(a,d)$ by opposite supertrace terms because this will not change the super-trace free condition. In super-twistor space this mean that we can add a $p=1$ term to the $n=2$ SCKT without losing super-tracelessness.
\subsection{$D=6$}

For $D=6$ the super-conformal groups are $OSp(8|N)$, for various numbers of supersymmetries $N$. This acts linearly on the super-twistor space $\bbC^{8|2N}$, but although this space does have an orthosymplectic invariant under the super-conformal group, it is not a straightforward extension of super-Minkowski space, so that the situation it is not a straightforward super-version of the $N=0$ case. Nevertheless the generalisation is not that difficult. An $n$th-rank super-conformal tensor is given by a super-tableau of the form \eq{6.2}, but where now each box corresponds to a super-index $\cA$ for a vector in $\bbC^{8|2N}$, and such that the tensor represented by such a tableau is traceless with respect to the orthosymplectic metric on this space. Thus the components of an $n$th-rank SKCT are given by a tensor $K$ of the form
\be
K^{\cA_1\cA_2,\cB_1\cB_2,\ldots}\ ,
\ee
having $n$ pairs of graded-antisymmetric indices, graded-symmetric under the interchange of pairs, and such that graded-antisymmetrisation over any three indices vanishes.  Furthermore, the trace with respect to the orthosymplectic metric on any pair of  indices vanishes. Such a tensor can be expanded out into tensors under $\go(8)\oplus \gs\gp(N)$, where the $\go(8)$ factor corresponds to the spacetime conformal algebra and the $\gs\gp(N)$ factor to the internal symmetry algebra, the index $i=1,\ldots 2N$ being considered as an odd index.

Reducibilty problems can also arise in $D=6$, starting at $n=2$. This tensor representation has the graded symmetries of the Riemann tensor, so that, when the super-traces are removed, one would expect to find a tensor with the graded symmetry properties of the Weyl tensor. However, one can show that it is not possible to remove the super-traceless Ricci tensor in $N=3$, so that we again have a reducibility problem. 

Note that the only problems of this type that occur have $N>2$ in $D=6$ and $N>4$ in D=4 (except for the algebra itself as we have discussed above), and are are therefore of limited interest in a physical context. This is because there are no non-trivial superconformal field theories that exceed these bounds.

\subsection{Algebras}

In the purely even case conformal Killing tensors define symmetries of the Laplacian \cite{Eastwood:2002su}, that is, linear differential operators $\cD$ that preserve the Laplacian in the sense that
\be
\D\cD=\d \D
\la{6.1}
\ee
where $\d$ is another linear differential operator. Clearly, $\cD$ maps solutions to Laplace's equation to other solutions. Each such $\cD$ has  a leading term given by a CKT, and this can be extended in a natural way to lower-order terms. Moreover, the product of two such symmetries defines a third (modulo the Laplacian) and so we get an algebra, known as the Eastwood algebra. In flat space, as we have seen, the components of any CKT are given by representations of the conformal algebra, so that this product can be described in Lie algebraic terms. Denoting the conformal Lie algebra by $\gg$, we can describe the Eastwood algebra as the tensor algebra of $\gg$ modulo its Joseph ideal \cite{Joseph:76} which is generated by
\be
X\otimes Y-X\circledcirc Y-\half[X,Y] - c <X,Y>
\ee
where $X,Y\in\gg$, $<X,Y>$ is the Killing form on $\gg$ with $c$ a constant and $X\circledcirc Y$ denotes the Cartan product which is the highest weight representation contained in the product. In the conformal case with $\gg=\go(2,D-2)$ it is just given by the two-row two-column tableau which is also traceless. The Cartan product extends to arbitrary CKT representations: for an $n$th rank and $m$th rank tensors we simply get the traceless two-row tableau with $n+m$ columns. In other words, the antisymmetric terms in the product are determined by the Lie bracket in $\gg$ while the symmetric terms are determined by the trace in $\gg$ and the symmetrised traceless product. The leading (symmetric)  term in this product is just given by the highest weight in the decomposition of the representations involved, while the first (antisymmetric) term coincides with the Schouten-Nijenhuis bracket \cite{Eastwood:2002su}.

Alternatively \cite{Eastwood:2002su}, the algebra can be viewed as the
the universal enveloping algebra, $\cU_{\gg}$, of $\gg:=\go(2,D-2)$, modulo the two-sided ideal generated by
\be
XY+YX-2X\circledcirc Y -2c <X,Y>\ .
\la{6.3}
\ee
The constant $c$ in the conformal case in $D$ dimensions is $\frac{D-2}{4(D+1)}$. This algebraic definition extends to simple complex Lie algebras, see \cite{Bravermann:98,Eastwood:2007}. For the classical cases, bar $\gs\gl(2)$, there is a unique value of the constant $c$ such that the quotient algebras are infinite-dimensional.

In the supersymmetric case, as we have mentioned, super Minkowski spaces cannot be presented in terms of  higher-dimensional superspaces of a similar type carrying a linear action of the appropriate superconformal group, but it should be feasible to generalise the purely Lie algebraic approach, provided that due care is taken with reducibility. However, in the context of super-Euclidean spaces equipped with ortho-symplectic metrics, Eastwood's Laplacian symmetry formalism can be extended to a natural super-Laplacian more or less straightforwardly \cite{Coul:2012}. Although this is not what we are directly interested in, it is nevertheless the case that the algebraic structures that arise in this situation (and studied in \cite{Coul:2012}) should be related to superconformal symmetry for $D=3,6$.  In the context of super Minkowski space a different notion of a super-Laplacian as a set of differential operators is more natural. We shall postpone a discussion of this and the related algebraic structures to a follow-up paper \cite{HL}.

In addition, super Joseph ideals have featured in a series of papers on quasi-conformal methods \cite{Govil:2013uta,Govil:2014uwa,Fernando:2015tiu}, although from a somehwat different point of view to ours. This work also has applications to higher spin and AdS/CFT, which we briefly discuss in the next section.

\section{Comments on higher spin}

Higher-spin fields were originally introduced by Fronsdal \cite{Fronsdal:1978rb} and the theory of them was subsequently developed in a series of papers by Vasiliev, see, for example, \cite{Fradkin:1987ah,Vasiliev:1999ba}. Further developments have included the incorporation of supersymmetry, see \eg \cite{Gates:1996my,Sezgin:1998gg,Sezgin:2000hr,Sezgin:2012ag}. The natural setting for higher-spin gauge theories is anti-de Sitter spacetime and the algebraic structures that arise reflect this; in particular, the fact that the symmetry algebra of AdS spacetime is  isomorphic to the conformal symmetry algebra of Minkowski space in one dimension lower implies that the AdS/CFT correspondence \cite{Maldacena:1997re,Gubser:1998bc,Witten:1998qj} is relevant in this higher-spin context \cite{Sezgin:2002rt,Klebanov:2002ja}. 

The AdS/CFT correspondence can be used to make the connection with CKTs. A Fronsdal higher-spin gauge field in the bulk gives rise to a related field on the boundary which couples naturally to a conserved, symmetric, traceless  tensor current of rank $(n+1)$, say. Such a current can be contracted on $n$ of its indices with an $n$th rank conformal Killing tensor, so that for each such current there is a conserved vector \cite{Mikhailov:2002bp}. These in turn can couple to fields in the bulk so that we arrive at a set of one-form gauge fields, parametrised by the CKTs,  in AdS. In this picture the underlying boundary fields are free, but one can attempt to introduce interactions in the bulk by including non-abelian terms in the field strengths. In order to do this one has to introduce algebraic structures. The question of the uniqueness of higher-spin theories in AdS and boundary CFTs has been clarified for the case of three-dimensional boundaries in \cite{Maldacena:2012sf}, and a more recent general study was presented in \cite{Boulanger:2013zza} where it was shown that in most dimensions of the bulk this algebra is unique and determined by the Eastwood algebra discussed above. (In this context it is known as the Eastwood-Vasiliev algebra). 

In \cite{Bianchi:2006ti} it was argued that the massless limit of IIB string theory on $AdS_5 \xz S^5$ would correspond, via the AdS/CFT correspondence, to a free $N=4$ super Yang-Mills theory on the boundary, in the sense that massless string gauge fields should couple to currents in the free $N=4$ theory on the boundary in the way described above. In addition, it was shown in \cite{Bianchi:2005ze} that the currents on the boundary can be explicitly constructed rather straightforwardly in $D=4, N=4$ analytic superspace. These currents are constructed in terms of two free field-strength superfields with linear combinations of analytic superspace derivatives acting on them in a similar manner to the construction of currents in terms of free scalar fields given in \cite{Mikhailov:2002bp}.\footnote{A study of higher spin superfields and $N=2$ supersymmetry in $N=1,D=4$ superspaces was given in \cite{Gates:1996my}, where a connection to superstrings was also conjectured.}

\section{Concluding remarks}

In this paper we have given a general definition of a superconformal Killing tensor in curved superspace subject to the constraint \eq{2.1} (as well as some conventional ones). Since \eq{2.1} is invariant under local scale transformations this leads naturally to placing the emphasis on superconformal Killing tensors. The most significant point of the general discussion is that SCKTs can be defined as purely even traceless symmetric tensors subject to the constraint that the smaller of the two spin representations that arise when one differentiates with respect to the spinorial covariant derivative should be set to zero. We then discussed these objects in the context of superparticles and in various flat superspaces as well as analytic superspaces and super twistor spaces. In the last case we were able to exhibit SCKTs explicitly as tensors carrying irreducible representations of the appropriate superconformal groups. We also indicated how this should lead to algebraic structures on the space of all SCKTs for a given dimension and number of supersymmetries. These symmetry algebras can then be related, via the AdS/CFT correspondence, to higher-spin structures in the bulk. As we mentioned in the text, we shall develop some of these ideas further in \cite{HL}; in particular, we shall study super-Laplacians and their symmetries.

Although we define SCKTs in curved superspace, the detailed discussion of components, etc, we have given here is limited to flat superspaces.  It should be mentioned that ordinary KTs in spaces of constant curvature are all reducible, \ie generated by KVs, as detailed in \cite{Weir}, \cite{Rani:2003br}, \cite{Thompson}, \cite{Mikhailov:2002bp}. It seems likely, but remains to be proven,  that a similar theorem holds for superspaces of this type.

Apart from the higher-spin connection, the other main application of (C)KTs is to situations where a given spacetime admits an irreducible KT. Some examples of this include the case of a fourth-order tensor found in the context of two-particle mechanics \cite{Cariglia:2014dfa}, a particle with worldline supersymmetry \cite{Lischewski:2014ffa} and the Perry-Myers black hole \cite{Myers:1986un}. When irreducible KTs exist, they may be used to separate coordinates in the Hamilton Jacobi equation. See \cite{Chervonyi:2015ima} for a recent discussion of this in the context of string theory.

It would be an interesting challenge to see if one could find any non-trivial higher-rank SKTs in non-trivial supergravity solutions formulated in superspace.

\section*{Acknowledgements}
UL gratefully acknowledges the hospitality of the theory group at Imperial College, London, as well as partial financial support by the Swedish Research Council through VR grant 621-2013-4245.

\appendix
\section{Appendix: Supersymmetric SN bracket}

Here we demonstrate that the function on the left-hand side of \eq{3.13} does indeed satisfy \eq{2.12}. We do this for the case $n=1,m=2$, but it would be easy to extend the proof to the general case. We want to show that
\be
M^{ab}:=-[K,L]^{ab}=2 L^{\{a|c|} \nab_c K^{b\}} -K^c\nab_c L^{ab}-2iK^\c (\C^{\{a})_{\c\d} L^{b\}\d}
\ee
satisfies
\be
\nab_\a M^{ab}=2i(\C^{\{a})_{\a\b} M^{b\}\b}
\la{B1}
\ee
for some (gamma-traceless) $M^{a\b}$. Since $K$ and $L$ are respectively a SCKV and a second-rank SCKT, we have
\begin{align}
\nab_\a K^a&=i(\C^a)_{\a\b} K^\b\nn\w1
\nab_\a K^\b+K^D T_{D\a}{}^\b&=\tilde S(K)_\a{}^\b\nn\w1
\nab_\a L^{ab}&=2i(\C^{\{a})_{\a\b} L^{b\}\b}\nn\w1
\nab_\a L^{a\b}+L^{aD} T_{D\a}{}^\b +L^{\b D} T_{D\a}{}^a&=\tilde S(L)_\a{}^{\b,a}
\label{B2}
\end{align}
where the $\tilde S$ functions are the generalised Lorentz and scale functions associated with $K$ and $L$ (called $\tilde L$ in section 2). 

When we apply $\nab_\a$ to \eq{B1} some terms are in the required form straight away, but others need some work. For example, applying $\nab_\a$ to $L^{ac}$ in the first term and using the third of \eq{B2} we get a term $2i(\C^a)_{\a\b} L^{c\b}\nab_c K^b$, which is fine, and there are two other terms like this. When the odd derivative hits the second factor in each of the first two terms it has to be taken past the even derivative so that it can act on $K$ or $L$. This gives rise to torsion and curvature terms. The torsion terms involve $T_{d\a}{}^\b$, but there are also terms like this coming from differentiating the  third term in $M^{ab}$. These come from the torsion terms in \eq{B2}. It turns out that the sum of all such terms vanish. There are also two dimension one-half torsion terms coming from differentiating the third term in $M$, which are contracted with the dimension-zero torsion. On using the dimension one-half Bianchi identity, we find that these give a term $2i(\C)_{\a\b} T_{\c\d}{}^\b K^\c L^{b\d}$, which is of the required form. The curvature terms, coming from the first two terms in $M$, sum up
 to
\be
4 L^{\{a|c} K^{d |} R_{\a[c,d]}{}^{b\}}=-2i (\C^{\{a})_{\a\b} L^{b\}c} K^d T_{cd}{}^\b\ ,
\la{B3}
\ee
where use was made of a dimension-three-halves Bianchi identity. In order to show that the final terms can be written in the desired form we have to make use of the identities that relate $\tilde S(K)_\a{}^\b$ and $\tilde S(L)_\a{}^{\b,c}$ (which arise from differentiating the third term and using \eq{B2}) to $\tilde S(K)_c{}^d$  and $\tilde S(L)_c{}^{d,e}$, which arise from the other terms with even derivatives on $K,L$. The final result is
\begin{align}
M^{a\b}&=L^{b\b}\nab_b K^a + L^{bc} \nab_c K^\b-K^b\nab_b L^{a\b}-L^{ac}K^d T_{cd}{}^\b-K^\c L^{a\d} T_{\c\d}{}^\b\nn\w1
&\ \ \ + L^{a\c} \tilde S(K)_{\c}{}^\b -K^\c \tilde S(L) _\c{}^{\b,a}\ ,
\la{B4}
\end{align}
where we assume that the gamma-trace has been removed.

An alternative way of proving this result is to differentiate the Poisson bracket of two functions $K,L$ of the type discussed in \eq{3.2c} with respect to time, and then make careful use of the equations of motion to show that $(K,L)$ is itself conserved. This method can be applied to SCKTs of arbitrary rank and therefore establishes the general result.


\end{document}